\documentclass[structabstract]{aa}  
\usepackage{graphicx}
\usepackage{amsmath,amssymb}
\usepackage[applemac]{inputenc}
\usepackage[modulo, switch]{lineno}
\usepackage{xcolor}
\usepackage{txfonts}
\usepackage{natbib}
\bibpunct{(}{)}{;}{a}{}{,} % to follow the A&A style
\newcommand{\kepler}{\textit{Kepler} }

\newcommand{\Dnu}{\Delta \nu}
\newcommand{\Dnunu}{\Delta\nu_{\ell} \left(\nu\right)}
\newcommand{\Dnunua}{\Delta_2\nu_{\ell} \left(\nu\right)}
\newcommand{\numax}{\nu_\mathrm{max}}

\newcommand{\muhz}{$\mu$Hz}

\newcommand{\diamonds}{\textsc{D\large{iamonds}}}

\newcommand{\kic}{KIC~12008916}
\newcommand{\he}{He\textsc{\,\large ii}}
\def\be{\begin{equation}}
\def\ee{\end{equation}}    
\def\ba{\begin{eqnarray}}
\def\ea{\end{eqnarray}}

\begin{document}

\title{High-precision acoustic helium signatures in 18 low-mass low-luminosity red giants}
\subtitle{Analysis from more than four years of \kepler observations}
\author{E. Corsaro\inst{1,2,3},
J. De Ridder\inst{3},
R. A. Garc\'{i}a\inst{2}
          }
\offprints{Enrico Corsaro\\ \email{enrico.corsaro@cea.fr}}

\institute{Instituto de Astrof\'{i}sica de Canarias, E-38205 -- Universidad de La Laguna, Departamento de Astrof\'{i}sica, E-38206, La Laguna, Tenerife, Spain
\and Laboratoire AIM, CEA/DSM -- CNRS -- Univ. Paris Diderot -- IRFU/SAp, Centre de Saclay, 91191 Gif-sur-Yvette Cedex, France
\and Instituut voor Sterrenkunde, KU Leuven, Celestijnenlaan 200D, B-3001 Leuven, Belgium
}

   \date{Received 19 February 2015 / Accepted 13 April 2015}

\abstract
{High-precision frequencies of acoustic modes in red giant stars are now available thanks to the long observing length and high-quality of the light curves provided by the NASA \kepler mission, thus allowing to probe the interior of evolved cool low-mass stars with unprecedented level of detail.} 
{We characterize the acoustic signature of the helium second ionization zone in a sample of 18 low-mass low-luminosity red giants by exploiting new mode frequency measurements derived from more than four years of \kepler observations.}
{We analyze the second frequency differences of radial acoustic modes in all the stars of the sample by using the Bayesian code \diamonds.}
{We find clear acoustic glitches due to the signature of helium second ionization in all the stars of the sample. We measure the acoustic depth and the characteristic width of the acoustic glitches with a precision level on average around $\sim$2\,\% and $\sim$8\,\%, respectively. We find good agreement with theoretical predictions and existing measurements from the literature. Lastly, we derive the amplitude of the glitch signal at $\numax$ for the second differences and for the frequencies with an average precision of $\sim$6\,\%, obtaining values in the range 0.14-0.24\,\muhz, and 0.08-0.33\,\muhz, respectively, which can be used to investigate the helium abundance in the stars.}
{}

\titlerunning{High-precision acoustic helium signatures in 18 low-mass low-luminosity red giants}
      \authorrunning{E. Corsaro et al.}
\maketitle
%
%==========================================================================
\section{Introduction}
\label{sec:intro}
The so-called acoustic glitches are regions of sharp-structure variation located in the interior of the stars and caused by the presence of a change in the energy transport from radiative to convective, by a rapid variation in the chemical composition, or by ionization zones of chemical elements such as hydrogen and helium. As originally predicted for the Sun \citep[e.g.][]{Vorontsov88,Gough90}, these regions produce tiny and regular variations in the frequency of the acoustic ($p$) modes that can be detected by direct measurement of the characteristic large frequency separation, namely the frequency separation between modes having the same angular degree. 

By studying the glitch signature in the Sun, it was possible to measure the acoustic position of the base of the convective zone and of the helium second ionization (\he) zone, as well as to provide estimates of the helium abundance in the envelope, and the extent of the overshooting \citep[e.g.][]{Basu95helium,Basu97,Monteiro05,CD11overshoot}. Expected also for distant stars \citep[e.g.][]{Monteiro2000,Mazumdar01,Ballot04}, thanks to the advent of the CoRoT \citep{Baglin06} and \kepler space missions \citep{Borucki10,Koch10}, which have released an outstanding amount of high-quality photometric observations for thousands of stars, these frequency shifts have been discovered and analyzed in many low-mass main-sequence, sub giant and red giant stars (RGs)  \citep{Mosser10,Miglio10,Mazumdar12,Mazumdar14,Verma14}, allowing to constrain both the position of the base of the convective zone and that of the \he\,\,zone in main-sequence and sub-giant stars, and the position of the \he\,\,zone in the case of the red giants.

The asteroseismology of red giant stars, in particular, has brought several important breakthroughs in the stellar physics of low-mass stars in the latest years \citep[e.g.][]{Beck11Science,Mosser11mixed,Bedding11Nature,Beck12Nature,Deheuvels12}. The characterization of the glitch signatures is able to provide tighter constraints on the chemical composition and the internal stratification of the star, and potentially allows to retrieve helium abundances in distant stars, essential for population studies \citep[e.g. see][hereafter B14, and references therein]{Broomhall14}. More recent studies focusing toward these evolved cool stars have analyzed the glitches due to the \he\,\,zone for an ensemble of more than a hundred targets observed by \kepler \citep{Vrard14}, and thoroughly investigated the properties of the signature from a theoretical point of view (B14, see also \citealt{Dalsgaard14} for more discussion). 

The recent availability of \kepler datasets spanning more than four years of nearly continuous observations, coupled with the development of new computational advances in asteroseismic data analysis \citep[e.g. see][hereafter C15]{Corsaro14,Corsaro15cat}, enables the study of the acoustic glitch signatures in red giants with an unprecedented level of detail.

In this paper, we report on the evidence of clear acoustic glitch signatures due to the \he\,\,zone in the sample of 19 red giants recently investigated by C15, hence we fully characterize the oscillatory signal by means of a Bayesian approach.

\section{Data analysis} 
\label{sec:analysis}
As noticed by C15, the low-mass low-luminosity red giants (LRGs) are well suited candidates to test stellar structure models and stellar evolution theory. The less-evolved stage in the red giant branch (RGB) of the stellar evolution for the LRGs, implies the highest frequency of maximum power $\numax$ for a red giant (between 100 and 200\,\muhz), hence a broader power excess caused by the oscillations, and consequently a larger number of radial orders observed (in general between six and nine). By having a larger number of high signal-to-noise ratio p-mode frequencies available, one is thus able to constrain the signature of the glitches more efficiently.

In this work we analyze the sample of LRGs studied by C15, who peak bagged their full oscillation spectrum using \kepler observations from Q0 till Q17.1, a total of $\sim$1470 days, with a frequency resolution of $\delta \nu_\mathrm{bin}\simeq 0.008\,$\muhz. The stars have $\numax$ values ranging from 110 to $\sim190$\,\muhz\,\,and masses in the interval $1$-$2\,M_\odot$. We adopt the high-precision individual frequency measurements from C15, and follow the theoretical approach by B14.

In the present analysis we refer to the first (frequency) difference, as the large frequency separation of a given angular degree, as a function of the frequency in the power spectral density of the star, $\Dnunu$. For a radial order $n$, $\Dnunu$ is thus defined as 
\begin{equation}
\Delta\nu_{n,\ell} \equiv \nu_{n+1,\ell} - \nu_{n,\ell} \, .
\label{eq:first_difference}
\end{equation}
where $\nu_{n,\ell}$ is the central frequency of the mode with angular degree $\ell$ and radial order $n$.
In addition, we compute the second (frequency) difference \citep[see e.g.][]{Gough90}, $\Dnunua$, defined for a single radial order as 
\begin{equation}
\begin{split}
\Delta_2\nu_{n,\ell} &\equiv \nu_{n+1,\ell} - 2 \nu_{n,\ell} + \nu_{n-1, \ell} \\
&= \Dnu_{n,\ell} - \Dnu_{n-1,\ell} \, .
\label{eq:second_difference}
\end{split}
\end{equation}
We fit the acoustic glitch signatures with the model introduced by \cite{Houdek07}, and used by B14 for RGs, defined as
\begin{equation}
\begin{split}
\Delta_2 \omega_{n,\ell} &= A \omega_{n,\ell} \exp( -2 b^2 \omega^2_{n,\ell}) \cos{\left[ 2 \left( \tau_\mathrm{He\,II} \omega_{n,\ell} + \phi \right) \right]} + c \, ,
\label{eq:glitch_model}
\end{split}
\end{equation}
with $\omega_{n,\ell} \equiv 2\pi \nu_{n,\ell}$ and $\Delta_2 \omega_{n,\ell} \equiv 2\pi \Delta_2 \nu_{n,\ell}$, $A$ a dimensionless amplitude of the signature signal, $\tau_\mathrm{He}$ the acoustic depth of the \he\,\,zone, $b$ its characteristic width, $\phi$ and $c$ a constant phase shift and offset, respectively, of the oscillatory signal. Following the arguments discussed by B14, we apply the fit to the second differences only because they are less prone to additional varying components such as hydrogen ionization and non-adiabatic processes, and to the general frequency dependence of the large separation caused by the development of the second-order term of the asymptotic relation \citep{Mosser11universal}. The second differences are at the same time available in a reasonably high number of measurements (two less than the total number of modes obtained for a given angular degree), still allowing to constrain the model parameters without leading to degeneracies in the solutions. 

Despite the possible presence of the oscillatory component in modes of angular degree $\ell > 0$, we point out that only radial mode frequencies are used for the final fit. The reason behind this choice mainly is the need to exploit pure p-mode character oscillations (see also B14), which in the case of RGs are only available in the form of $\ell = 0$ modes. This is because the coupling occurring between $p$ modes of angular degree $\ell > 0$ and $g$ modes arising from the radiative interior can hamper the asymptotic behavior of the corresponding modes by producing so-called mixed modes, whose frequencies deviate from the expected position of a pure $p$ mode oscillation \citep[e.g.][]{Beck11Science}.

We perform all the fits following a Bayesian approach done by means of \diamonds\,\,\citep{Corsaro14}, hence exploiting a nested sampling Monte Carlo method to perform the inference and estimate the free parameters of the model given by Eq.~(\ref{eq:glitch_model}) from their individual marginal probability distributions (see \citealt{Corsaro14} for more details). The configuring parameters of \diamonds\,\,(following the definitions by \citealt{Corsaro14}) used for all the computations are: initial enlargement fraction $1.0 \leq f_0 \leq 1.7$, shrinking rate $\alpha = 0.02$-$0.03$, number of live points $N_\mathrm{live} = 1000$, number of clusters $1 \leq N_\mathrm{clust} \leq 4$, number of total drawing attempts $M_\mathrm{attempts} = 10^4$, number of nested iterations before the first clustering $M_\mathrm{init} = 1000$, and number of nested iterations with the same clustering $M_\mathrm{same} = 50$.

For this analysis we adopt a normal likelihood function, as that used by \cite{Corsaro13}, which takes into account the uncertainties, with correlations included, on the measurements of the second differences. This assumes that the residuals arising from the difference between predicted and measured second differences are Gaussian distributed. Concerning the set up of priors, since we do not have initial guesses available from the literature for the given stars, we use uniform (i.e. flat) prior probability distributions for all the free parameters of Eq.~(\ref{eq:glitch_model}), with lower and upper boundaries for each parameter range obtained by comparison with existing measurements of the acoustic depths derived by \cite{Miglio10}, \cite{Mazumdar14}, and the theoretical results by B14 in the observed range of $\numax$. The choice of uniform priors also yields a faster computation with \diamonds, as already discussed by \cite{Corsaro14} and C15.

Following the discussion by \cite{Ballot04} and B14, we compute the acoustic radius of the \he\,\,zones, $t_\mathrm{He\,\,II}$, since it represents a quasi unbiased estimator of the acoustic position of the glitch. This is done by using the mean large frequency separation, $\langle \Dnu \rangle$, obtained from the radial mode frequencies provided by C15, giving the total acoustic radius of the star, $T = (2 \langle \Dnu \rangle)^{-1}$, hence the acoustic radius of the \he\,\,zone, $t_\mathrm{He\,\,II} \equiv T - \tau_\mathrm{He\,\,II}$. 

Lastly, to provide measurements that can be used to model the helium abundance in the envelope of the stars, following B14 we extract the amplitude of the signal at $\numax$ from Eq.~(\ref{eq:glitch_model}), obtaining
\begin{equation}
A_\mathrm{max} = A \numax \exp{(-2 b^2 \omega^2_\mathrm{max})} \, ,
\label{eq:amplitude}
\end{equation}
with $\omega_\mathrm{max} = 2 \pi \numax$, and $\numax$ derived from the background fit done by C15. Following \cite{Verma14}, we also derive the amplitude of the signal in the frequencies, $A_\mathrm{He}$, given as
\begin{equation}
A_\mathrm{He} = \frac{A_\mathrm{max}}{4 \cos^2 \left[ 2 \pi \tau_\mathrm{He\,II} \langle \Dnu \rangle \right] } \, ,
\label{eq:amp_he}
\end{equation}
where $\langle \Dnu \rangle$ is the same mean large frequency separation used to calculate $t_\mathrm{He\,\,II}$, and $\tau_\mathrm{He\,\,II}$ is the same acoustic depth used in Eq.~(\ref{eq:glitch_model}). The adoption of $A_\mathrm{He}$ to retrieve the helium content in the envelope is preferred since this value is not influenced by the location of the glitch \citep[see e.g.][]{Mazumdar01,Mazumdar14,Verma14}. For clarity to the reader, we stress that $A_\mathrm{max}$ is derived from a Bayesian approach by using the same sampling of the posterior probability distribution obtained by \diamonds\,\,for the free parameters of the glitch model (see also \citealt{Corsaro14}, Fig. 7, for an analogous case presenting the sampling from \diamonds\,\,for a combination of different inferred parameters). The parameters $t_\mathrm{He\,\,II}$, and $A_\mathrm{He}$, simply follow from their definitions presented above, by using both the value $\langle \Dnu \rangle$ computed from the radial mode frequencies presented by C15 for each star of our sample, and the estimated model parameters of Eq.~(\ref{eq:glitch_model}) (see Sect.~\ref{sec:results} for more details).

\section{Results}
\label{sec:results}
The results for $\Dnunu$, and $\Dnunua$ for the star \kic\,\,are shown in Fig.~\ref{fig:glitch} (top and bottom panel, respectively), and can be found in Appendix~\ref{sec:glitch_results} for all the other LRGs, together with the tables with the individual measurements of the radial angular frequencies $\omega_{n,0}$, and of the corresponding second angular frequency differences $\Delta_2 \omega_{n,0} \left(\nu\right)$ used in this work. We discarded KIC~10123207 from the fit because of the low number of available measurements (four second differences only, one less than the minimum required to fit the model given by Eq.~\ref{eq:glitch_model}). To help the reader visualize the presence of the oscillatory signal in the first differences of the angular degrees $\ell = 0, 2, 3$ and in the second differences of the angular degrees $\ell = 2, 3$, we included low-degree (3-4) polynomial fits. The 1-$\sigma$ uncertainties on the first and second differences derived from a standard error propagation of the uncertainties of the individual mode frequencies, following Eq.~(\ref{eq:first_difference}) and Eq.~(\ref{eq:second_difference}), respectively, are overlaid in each plot, though they are not visible in most of the cases because they are smaller than the size of the symbols used for the measurements. The uncertainties on all the measurements are listed in the corresponding tables in Appendix~\ref{sec:glitch_results} for completeness. 

\begin{figure}
   \centering
   \includegraphics[width=9.0cm]{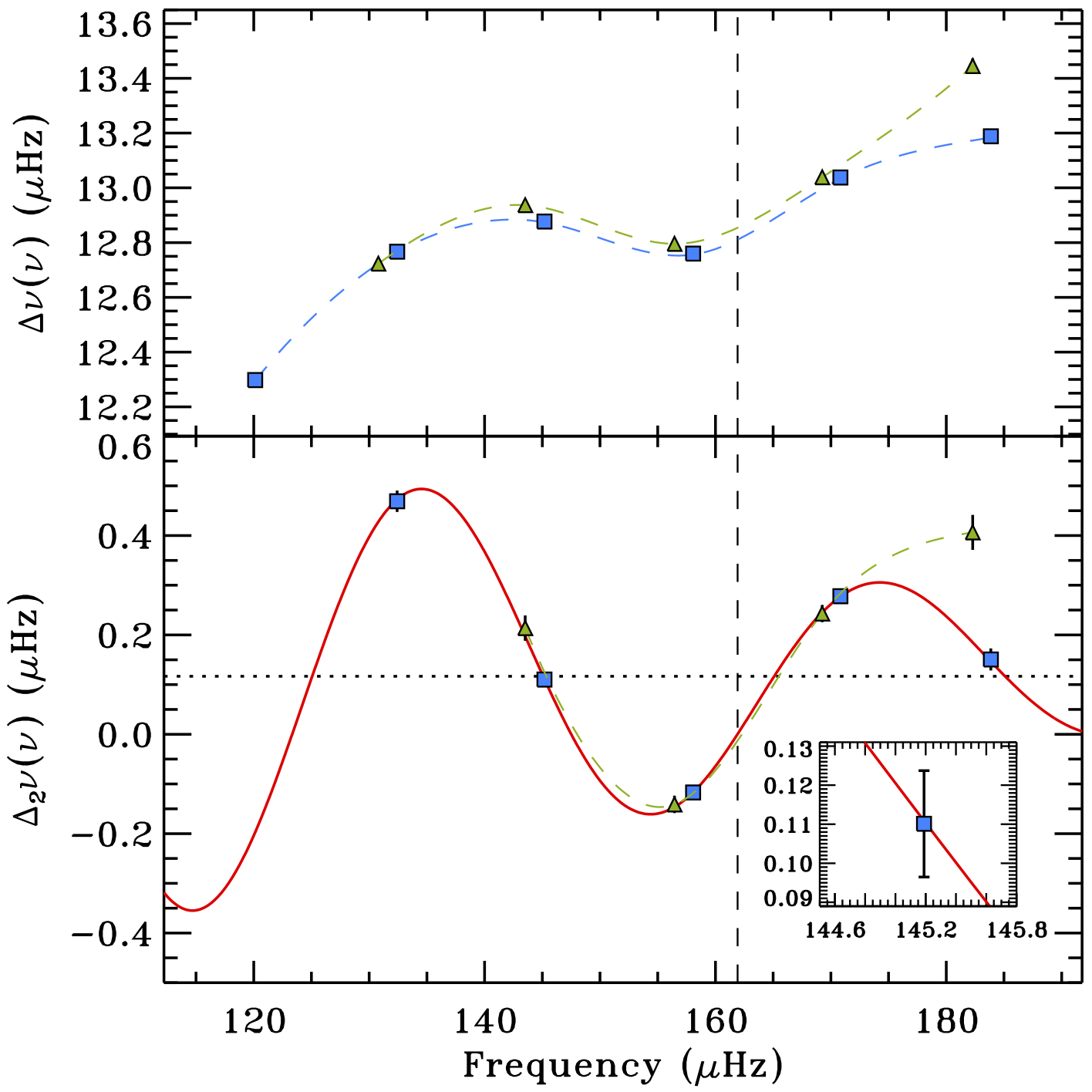}
      \caption{Acoustic glitches for \kic. \textit{Top panel}: the first difference, $\Delta\nu \left( \nu \right)$, Eq.~(\ref{eq:first_difference}). Blue squares represent values computed from $\ell = 0$ modes, while green triangles from $\ell = 2$ modes, with polynomial fits (dashed lines with same color as symbols) overlaid to visualize the oscillatory trend. The dashed vertical line marks the position of $\numax$ for a reference. \textit{Bottom panel}: the second difference $\Delta_2 \nu \left( \nu \right)$, Eq.~(\ref{eq:second_difference}), with same symbol description as for the top panel. The solid red line indicates the fit to the $\ell = 0$ measurements given by Eq.~(\ref{eq:glitch_model}) with the estimated parameters listed in Table~\ref{tab:glitch_parameters}, as derived by \diamonds. The horizontal dotted line marks the offset level $c/(2\pi)$, useful to visualize the amplitude of the signature. The inset shows a zoom in of one of the measurements to visualize the precision-level of the fit.}
    \label{fig:glitch}
\end{figure}

We find that all the stars analyzed have clear acoustic glitches due to the signature of the \he\,\,zone in $\ell = 0$ and 2 modes up to the second differences. We can see the presence of acoustic glitches also in $\ell = 3$ modes for most of the stars thanks to at least four different frequency measurements available. As mentioned in Sect.~\ref{sec:analysis}, we find that the measurements for modes having angular degrees $\ell = 2, 3$ often deviate from those of the radial oscillations (see e.g. Figs.~\ref{fig:8366239glitch}, \ref{fig:8718745glitch}, \ref{fig:9267654glitch}, and~\ref{fig:10200377glitch}). As also indicated by C15 for the case of the mode linewidths of the $\ell = 2$ modes, this different behavior relies on the presence of both mixed quadrupole modes and rotational split components. When using an individual Lorentzian profile to fit the frequency region containing the oscillation peak, either an $\ell = 2$ or 3 mode, as done by C15, the effects mentioned before can significantly change the measured frequency of the peak. A reliable treatment of the mixed modes and of the rotational split components for $\ell = 2,3$ modes is however difficult due to the high proximity of the individual peaks.

The model fit to the acoustic glitch signatures of \kic\,\,is shown in the bottom panel of Fig.~\ref{fig:glitch} for the case of $\Delta_2 \nu_{n,0} \left(\nu \right)$ (red line), and similarly for the other stars in Appendix~\ref{sec:glitch_results}. All the estimated parameters of Eq.~(\ref{eq:glitch_model}) are provided in Table~\ref{tab:glitch_parameters}. The inset in the bottom panel of Fig.~\ref{fig:glitch} provides a closer view of one of the measurements to visualize the precision-level achieved in the fit. In particular, we find that the model given by Eq.~(\ref{eq:glitch_model}) yields a remarkable fit quality for most of the stars, with average uncertainties of $\sim$2\% for $\tau_\mathrm{He\,\,II}$, and $\sim8$\,\% for $b$. Following the analysis presented by \cite{Corsaro13}, we have obtained the weighted Gaussian rms of the residuals, $\sigma_\mathrm{rms}$ (listed in Table~\ref{tab:glitch_parameters} as well). For its computation we have adopted the weights $w_i =\sigma_i^{-2}$, $\sigma_i$ being the uncertainties on the second frequency difference coming from those reported in Table~\ref{tab:2diff} and \ref{tab:2diff_2}. The quantity $\sigma_\mathrm{rms}$ provides additional information to the reader because it allows to compare the quality of the fits between different stars, and to relate the precision achieved on the individual fits to the given uncertainties of the data points. We note that for all the fits presented in this work, the values for $\sigma_\mathrm{rms}$ are remarkably small, ranging from $10^{-2}$ down to $10^{-3}$ \muhz\,in the best cases, thus reaching in many cases the same order of the precision-level obtained on the individual frequencies of the radial modes. For a reference to the reader, in Table~\ref{tab:glitch_parameters} we also provide the values for the total acoustic radius $T$, with its 1-$\sigma$ standard deviation, and the values of $\numax$ obtained by C15.

The stars KIC~8475025, KIC~9145955, KIC~10200377, and KIC~11913545, each show a component at high frequency that is not properly predicted by the adopted model. This mainly relies on some residual frequency dependence of the second differences that becomes more pronounced toward the wings of the region containing the oscillations. However, we note that the measurements at higher frequencies all have larger error-bars (up to 10 times) with respect to the others, because of the larger linewidths of the peaks occurring at high-frequency (see C15 for more details). The fits derived, except for KIC~8475025, are therefore not significantly affected by the measurements of the second differences falling at high-frequency, whereas they are almost entirely constrained by those close to $\numax$. This is inspected by refitting the glitch model without the highest-frequency measurement of the second difference (showing the deviating behavior), hence noticing that the new estimated free parameters of the model lay well within the uncertainties of those reported in Table~\ref{tab:glitch_parameters}. For KIC~8475025 however, we find that the fit is unstable due to the large deviation (more than 0.2\,\muhz) of the second difference measurement falling at the highest frequency (see Fig.~\ref{fig:8475025glitch}). This is because the measured oscillation frequency of the highest-frequency radial mode is likely affected by additional sources such as mixed modes and rotational split components arising from the neighbor $\ell = 2$ mode, which are enhanced by the large mode linewidths (see C15 for more details). To stabilize the fit for this star and provide estimates of the model parameters that are comparable to the other stars in the sample, we have therefore chosen to discard the last measurement of the second difference for this particular target. In the case of KIC~8366239, KIC~9267654, and KIC~10200377, the highest-frequency values are marked as not reliable (open symbols), according to the Bayesian peak significance test done by C15, although they were included in the fit since they don't produce any significant change in the results for the same reasons discussed above.

\begin{figure}
   \centering
   \includegraphics[width=9.0cm]{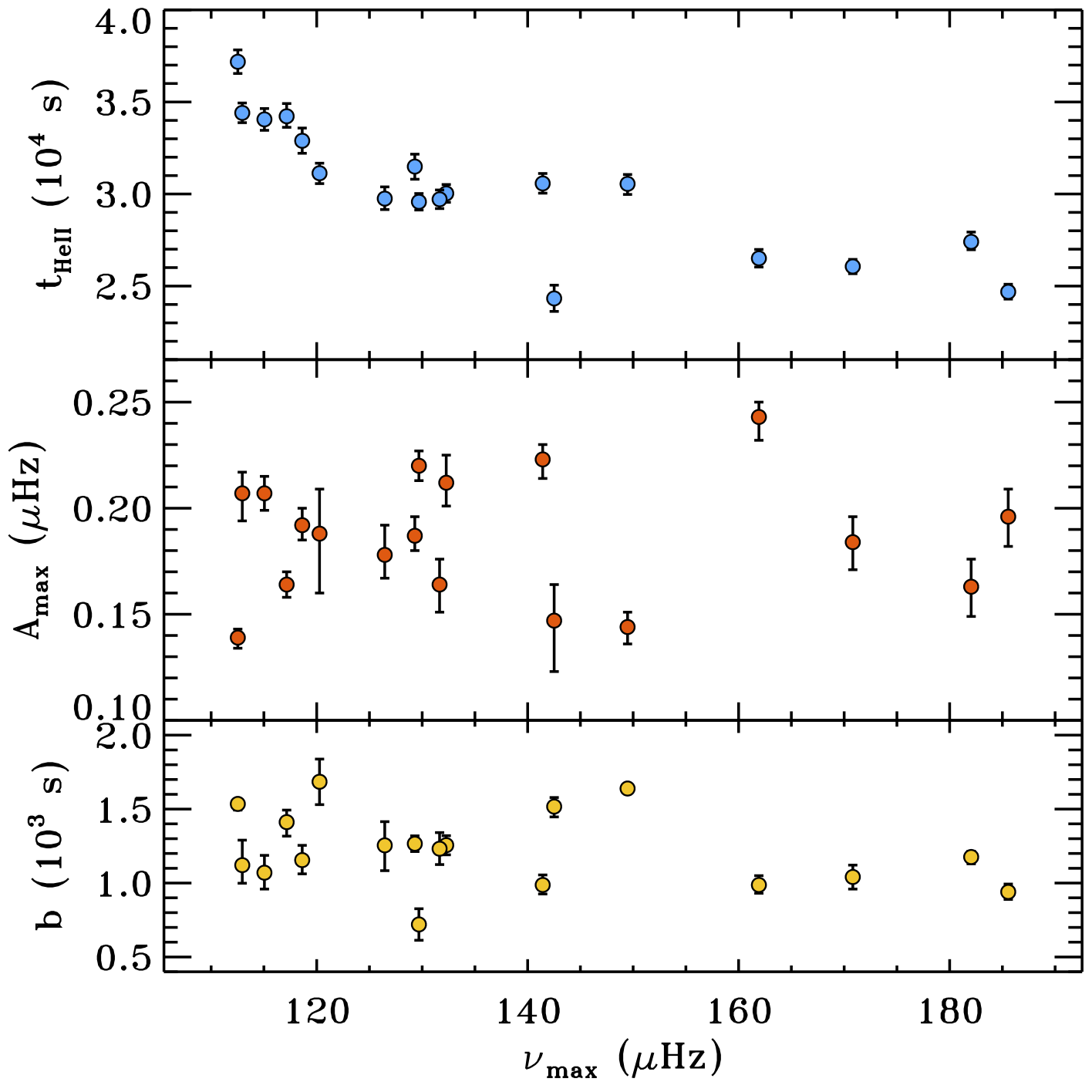}
      \caption{Acoustic radius of the \he\,\,zone (top panel) and corresponding amplitude of the oscillatory signal at $\numax$ (middle panel), and characteristic width $b$ (bottom panel), as a function of $\numax$ for all the stars of the sample. The 68\,\% Bayesian uncertainties listed in Table~\ref{tab:glitch_parameters} are overlaid for all the measurements.}
    \label{fig:helium}
\end{figure}

The measurements of the acoustic radius $t_\mathrm{He\,\,II}$, the amplitude of the signal $A_\mathrm{max}$ from Eq.~(\ref{eq:amplitude}), and the corresponding characteristic width $b$, are shown in Fig.~\ref{fig:helium} (top, middle and bottom panels, respectively) for all the stars of the sample. We note that while the model parameters (Eq.~\ref{eq:glitch_model}) and their corresponding 68\,\% Bayesian credible intervals are estimated by means of \diamonds\,\,(see Table~\ref{tab:glitch_parameters} and \citealt{Corsaro14} for more details on the derivation of the Bayesian uncertainties), the uncertainties for the additional parameters $t_\mathrm{He\,\,II}$, $A_\mathrm{max}$, and $A_\mathrm{He}$, were obtained in a subsequent step. In particular for $A_\mathrm{max}$, we have used the same sampling of the posterior probability distribution obtained by \diamonds, hence we have derived the median and the corresponding 68\,\% Bayesian credible intervals directly from the marginal probability distributions of $A_\mathrm{max}$. For $t_\mathrm{He\,\,II}$ and $A_\mathrm{He}$, the uncertainties follow from those of the acoustic depth through the definition of the acoustic radius, and by a rescaling of the uncertainties on $A_\mathrm{max}$ through Eq.~(\ref{eq:amp_he}), respectively. All the resulting values are listed in Table~\ref{tab:glitch_parameters} as well. We note that the precision obtained on our measurements of the acoustic radii of the \he\,\,zones is about 10 times higher than that obtained by \cite{Miglio10} using CoRoT data. In addition, all the values match those predicted by B14 along the entire range of $\numax$ investigated, showing a clear increasing trend toward lower $\numax$, as expected for more evolved stages of the evolution in the RGB. The derived amplitudes in frequency, $A_\mathrm{He}$, are within the range $0.08$-$0.33$\,\muhz, and are varying from star to star with uncertainties on average around $\sim$6\,\%, thus opening the possibility to study the He abundance by direct comparison with stellar models.

\begin{table*}
\caption{Median values with corresponding 68.3\,\% shortest credible intervals as derived by \diamonds\,\,for the free parameters of the model to fit the acoustic glitch signatures in the second differences, Eq.~(\ref{eq:glitch_model}), and for the acoustic radius of the \he\,zone, the total acoustic radius of the star, the amplitude at $\numax$ in the second difference and in frequency (Eqs.~\ref{eq:amplitude} and~\ref{eq:amp_he}, respectively).}             % title of Table
\centering                         
\begin{tabular}{l r c c c c c c c c c c}       
\hline\hline
\\[-8pt]
KIC ID & \multicolumn{1}{c}{$\tau_\mathrm{He\,II}$} & $A$ & $b$ & $\phi$ & $c$ & $t_\mathrm{He\,II}$ & $T$ & $A_\mathrm{max}$ & $A_\mathrm{He}$ & $\numax$ &$\sigma_\mathrm{rms}$ \\ [1pt]
 & \multicolumn{1}{c}{(s)} &  & (s) & (rad) & \multicolumn{1}{c}{(\muhz)} & (s) & (s) & (\muhz) & (\muhz) & (\muhz) & (\muhz) \\ [1pt]
\hline
\\[-8pt]
03744043 & $     13468_{-     272}^{+     258}$ & $    0.0131_{-    0.0015}^{+    0.0010}$ & $    1535_{-      42}^{+      27}$ & $         1.9_{-         0.2}^{+         0.2}$ & $        0.86_{-        0.03}^{+        0.03}$ & $       37183_{-         638}^{+         643} $ & $50652_{-584}^{+584}$ &$       0.139_{-       0.005}^{+       0.004}$ & $       0.077_{-       0.003}^{+       0.002}$ & 112.52 & $       0.007$ \\[1pt]
06117517 & $     18422_{-     263}^{+     303}$ & $    0.0403_{-    0.0214}^{+    0.0137}$ & $    1685_{-     155}^{+     153}$ & $         0.7_{-         0.2}^{+         0.2}$ & $        0.58_{-        0.05}^{+        0.04}$ & $       31128_{-         563}^{+         543} $ & $49551_{-475}^{+475}$ & $       0.188_{-       0.028}^{+       0.021}$ & $       0.306_{-       0.046}^{+       0.034}$ & 120.27 & $       0.017$ \\[1pt]
06144777 & $     16083_{-     230}^{+     228}$ & $    0.0034_{-    0.0007}^{+    0.0006}$ & $     720_{-     107}^{+     106}$ & $         1.4_{-         0.2}^{+         0.2}$ & $        0.67_{-        0.03}^{+        0.02}$ & $       29576_{-         448}^{+         449} $ & $45660_{-386}^{+386}$ & $       0.220_{-       0.007}^{+       0.007}$ & $       0.274_{-       0.009}^{+       0.009}$ & 129.69 & $       0.011$ \\[1pt]
07060732 & $     15886_{-     205}^{+     205}$ & $    0.0142_{-    0.0027}^{+    0.0023}$ & $    1256_{-      65}^{+      64}$ & $         1.4_{-         0.2}^{+         0.2}$ & $        0.65_{-        0.03}^{+        0.04}$ & $       30032_{-         482}^{+         482} $ & $45919_{-436}^{+436}$ & $       0.212_{-       0.011}^{+       0.013}$ & $       0.245_{-       0.013}^{+       0.015}$ & 132.29 & $       0.017$ \\[1pt]
07619745 & $     12231_{-     227}^{+     240}$ & $    0.0131_{-    0.0048}^{+    0.0034}$ & $    1041_{-      82}^{+      80}$ & $         1.5_{-         0.2}^{+         0.2}$ & $        0.63_{-        0.04}^{+        0.04}$ & $       26064_{-         391}^{+         383} $ & $38296_{-309}^{+309}$ & $       0.184_{-       0.013}^{+       0.012}$ & $       0.159_{-       0.011}^{+       0.010}$ & 170.82 & $       0.011$ \\[1pt]
08366239 & $     11826_{-     288}^{+     271}$ & $    0.0118_{-    0.0029}^{+    0.0022}$ & $     940_{-      51}^{+      53}$ & $         1.0_{-         0.3}^{+         0.3}$ & $        0.49_{-        0.05}^{+        0.06}$ & $       24686_{-         397}^{+         409} $ & $36513_{-291}^{+291}$ & $       0.196_{-       0.014}^{+       0.013}$ & $       0.177_{-       0.013}^{+       0.012}$ & 185.56 & $       0.020$ \\[1pt]
 08475025 & $     17796_{-     283}^{+     280}$ & $    0.0065_{-    0.0024}^{+    0.0017}$ & $    1121_{-     122}^{+     169}$ & $         2.0_{-         0.2}^{+         0.2}$ & $        0.61_{-        0.05}^{+        0.05}$ & $       34406_{-         533}^{+         535} $ & $52203_{-454}^{+454}$ & $       0.207_{-       0.013}^{+       0.010}$ &       $       0.225_{-       0.014}^{+       0.011}$ & 112.95 & $       0.023$ \\[1pt]
08718745 & $     12468_{-     415}^{+     434}$ & $    0.0120_{-    0.0021}^{+    0.0018}$ & $    1266_{-      53}^{+      53}$ & $         1.3_{-         0.3}^{+         0.3}$ & $        1.02_{-        0.05}^{+        0.04}$ & $       31490_{-         686}^{+         674} $ & $43958_{-531}^{+531}$ & $       0.187_{-       0.007}^{+       0.009}$ & $       0.118_{-       0.004}^{+       0.006}$ & 129.31 &$       0.012$ \\[1pt]
09145955 & $     15626_{-     274}^{+     271}$ & $    0.0100_{-    0.0034}^{+    0.0025}$ & $    1232_{-     107}^{+     109}$ & $         1.7_{-         0.2}^{+         0.2}$ & $        0.60_{-        0.04}^{+        0.04}$ & $       29712_{-         502}^{+         504} $ & $45339_{-424}^{+424}$ & $       0.164_{-       0.013}^{+       0.012}$ & $       0.186_{-       0.015}^{+       0.014}$ & 131.65 & $       0.015$ \\[1pt]
09267654 & $     15536_{-     430}^{+     410}$ & $    0.0072_{-    0.0018}^{+    0.0014}$ & $    1155_{-      93}^{+     100}$ & $         3.0_{-         0.3}^{+         0.3}$ & $        0.87_{-        0.04}^{+        0.04}$ & $       32891_{-         680}^{+         693} $ & $48427_{-544}^{+544}$ & $       0.192_{-       0.007}^{+       0.008}$ & $       0.169_{-       0.006}^{+       0.007}$ & 118.63 & $       0.018$ \\[1pt]
09475697 & $     16782_{-     296}^{+     304}$ & $    0.0060_{-    0.0016}^{+    0.0012}$ & $    1070_{-     111}^{+     117}$ & $         2.5_{-         0.2}^{+         0.2}$ & $        0.65_{-        0.04}^{+        0.03}$ & $       34052_{-         595}^{+         590} $ & $50835_{-512}^{+512}$ &  $       0.207_{-       0.008}^{+       0.008}$ & $       0.200_{-       0.008}^{+       0.008}$ & 115.05 & $       0.015$ \\[1pt]
09882316 & $      9232_{-     475}^{+     356}$ & $    0.0342_{-    0.0083}^{+    0.0071}$ & $    1176_{-      46}^{+      39}$ & $         4.1_{-         0.4}^{+         0.5}$ & $        0.73_{-        0.06}^{+        0.08}$ & $       27402_{-         429}^{+         532} $ & $36634_{-240}^{+240}$ &  $       0.163_{-       0.014}^{+       0.013}$ & $       0.083_{-       0.007}^{+       0.007}$ & 182.04 & $       0.011$ \\[1pt]
10200377 & $     15660_{-     645}^{+     626}$ &  $0.0415_{-    0.0088}^{+    0.0063}$ & $    1517_{-      70}^{+      61}$ & $         0.9_{-         0.6}^{+         0.5}$ & $        0.76_{-        0.03}^{+        0.04}$ & $       24327_{-         704}^{+         721} $ & $39988_{-323}^{+323} $  & $       0.147_{-       0.024}^{+       0.017}$ & $       0.330_{-       0.054}^{+       0.038}$ & 142.52 & $       0.029$ \\[1pt]
10257278 & $     10565_{-     353}^{+     441}$ & $    0.1106_{-    0.0155}^{+    0.0181}$ & $    1639_{-      29}^{+      32}$ & $         1.5_{-         0.4}^{+         0.3}$ & $        0.83_{-        0.05}^{+        0.05}$ & $       30551_{-         574}^{+         510} $ & $41116_{-369}^{+369}$ & $       0.144_{-       0.008}^{+       0.007}$ & $       0.075_{-       0.004}^{+       0.004}$ & 149.47 & $       0.007$ \\[1pt]
11353313 & $     16819_{-     447}^{+     388}$ & $    0.0104_{-    0.0053}^{+    0.0038}$ & $    1255_{-     171}^{+     160}$ & $         1.4_{-         0.3}^{+         0.3}$ & $        0.71_{-        0.05}^{+        0.04}$ & $       29749_{-         597}^{+         637} $ & $46568_{-455}^{+455} $ & $       0.178_{-       0.011}^{+       0.014}$ & $       0.249_{-       0.015}^{+       0.020}$ & 126.46 & $       0.006$ \\[1pt]
11913545 & $     15218_{-     482}^{+     338}$ & $    0.0122_{-    0.0034}^{+    0.0025}$ & $    1412_{-      95}^{+      81}$ & $         3.4_{-         0.3}^{+         0.3}$ & $        0.70_{-        0.03}^{+        0.03}$ & $       34222_{-         601}^{+         692} $ & $49440_{-498}^{+498}$ & $       0.164_{-       0.006}^{+       0.006}$ & $       0.127_{-       0.005}^{+       0.005}$ & 117.16 & $       0.005$ \\[1pt]
11968334 & $     13429_{-     313}^{+     294}$ & $    0.0074_{-    0.0015}^{+    0.0012}$ & $     987_{-      61}^{+      68}$ & $         2.6_{-         0.3}^{+         0.3}$ & $        0.72_{-        0.04}^{+        0.04}$ & $       30573_{-         529}^{+         540} $ & $44003_{-441}^{+441}$ & $       0.223_{-       0.009}^{+       0.007}$ & $       0.169_{-       0.007}^{+       0.005}$ & 141.43 & $       0.007$ \\[1pt]
12008916 & $     12492_{-     302}^{+     286}$ & $    0.0113_{-    0.0028}^{+    0.0023}$ & $     987_{-      56}^{+      63}$ & $         2.0_{-         0.3}^{+         0.3}$ & $        0.73_{-        0.05}^{+        0.04}$ & $       26505_{-         474}^{+         484} $ & $38998_{-379}^{+379}$ & $       0.243_{-       0.011}^{+       0.007}$ & $       0.212_{-       0.010}^{+       0.006}$ & 161.92 & $       0.013$ \\[1pt]\hline                                
\end{tabular}
\tablefoot{The parameters refer to the angular measurements of the second differences, $\Delta_2 \omega_{n,0} \left(\nu\right)$, and the corresponding frequencies $\omega_{n,0}$, of the radial modes only.}
\tablefoot{The last two columns provide the reference values for $\numax$ (provided by C15) and the weighted Gaussian rms of the residuals, as described in Sect.~\ref{sec:results}.}
\label{tab:glitch_parameters}
\end{table*}

\section{Conclusions}
\label{sec:conclusions}

By exploiting the set of individual mode frequencies extracted by C15 for a sample of 19 LRGs with a precision level up to $10^{-3}$\,\muhz, we computed the first differences, Eq.~(\ref{eq:first_difference}), and the second differences, Eq.~(\ref{eq:second_difference}), for presenting the results on the acoustic glitches of these stars. In this work, we have shown that the acoustic glitches are remarkably clear for all the red giants of the sample, and for both $\ell =0$ and $\ell = 2$ modes, up to the second difference (where five to seven different measurements are available for each star, except KIC~10123207 that instead has only four and was not considered in the analysis), with many cases involving $\ell = 3$ modes as well.

By adopting the model for the second differences introduced by \cite{Houdek07}, Eq.~(\ref{eq:glitch_model}), we have extracted the acoustic depth, the characteristic width and amplitude of the signal of all the \he\,\,zones of the stars in our sample (see Table~\ref{tab:glitch_parameters}) with an unprecedented level of detail for red giant stars (on average $\sim$2\,\% for the acoustic depths, $\sim$8\,\% for the characteristic widths, and $\sim$6\,\% for the amplitudes of the glitch signal in both second difference and frequency), improved by about one order of magnitude with respect to existing measurements of acoustic depths from the literature. These values, reflecting the high-precision obtained on the individual frequency measurements of the radial modes, confirm the theoretical predictions done by B14 in the same range of $\numax$. We also stress that the glitch model given by \cite{Houdek07} is able to predict the observations very exhaustively for most of the stars (well within the given uncertainties of the measurements for most of the data points available, as shown in Fig.~\ref{fig:glitch} and in the other figures in the Appendix). This is also supported by our computation of $\sigma_\mathrm{rms}$, listed in Table~\ref{tab:glitch_parameters}, which are on the same precision-level of that given by the measurements of the second difference for most of the stars analyzed.

Finally, the set of values for $A_\mathrm{max}$ and $A_\mathrm{He}$ derived in this work, the latter not being influenced by the position of the glitch, coupled with the high precision achieved, will be useful to investigate the helium content in the envelope of the stars, and possibly contribute to study scenarios of helium enrichment in low-mass stars \citep[e.g. see][and references therein]{Gratton12enrichment}.

\begin{acknowledgements}
E.C. is funded by the European Community's Seventh Framework Programme (FP7/2007-2013) under grant agreement n$^\circ$312844 (SPACEINN).
The research leading to these results has received funding from the European
Research Council under the European Community's Seventh Framework Programme
(FP7/2007--2013) ERC grant agreement n$^\circ$227224 (PROSPERITY), 
from the Fund for Scientific Research of Flanders (G.0728.11).
E.C. thanks A.-M. Broomhall and A. Miglio for useful discussions.
\end{acknowledgements}

\bibliographystyle{aa} % style aa.bst
%\bibliography{biblio} % your references

\Online

\appendix
\section{Results for the fitting of the acoustic signatures}
\begin{figure}
   \centering
   \includegraphics[width=9.0cm]{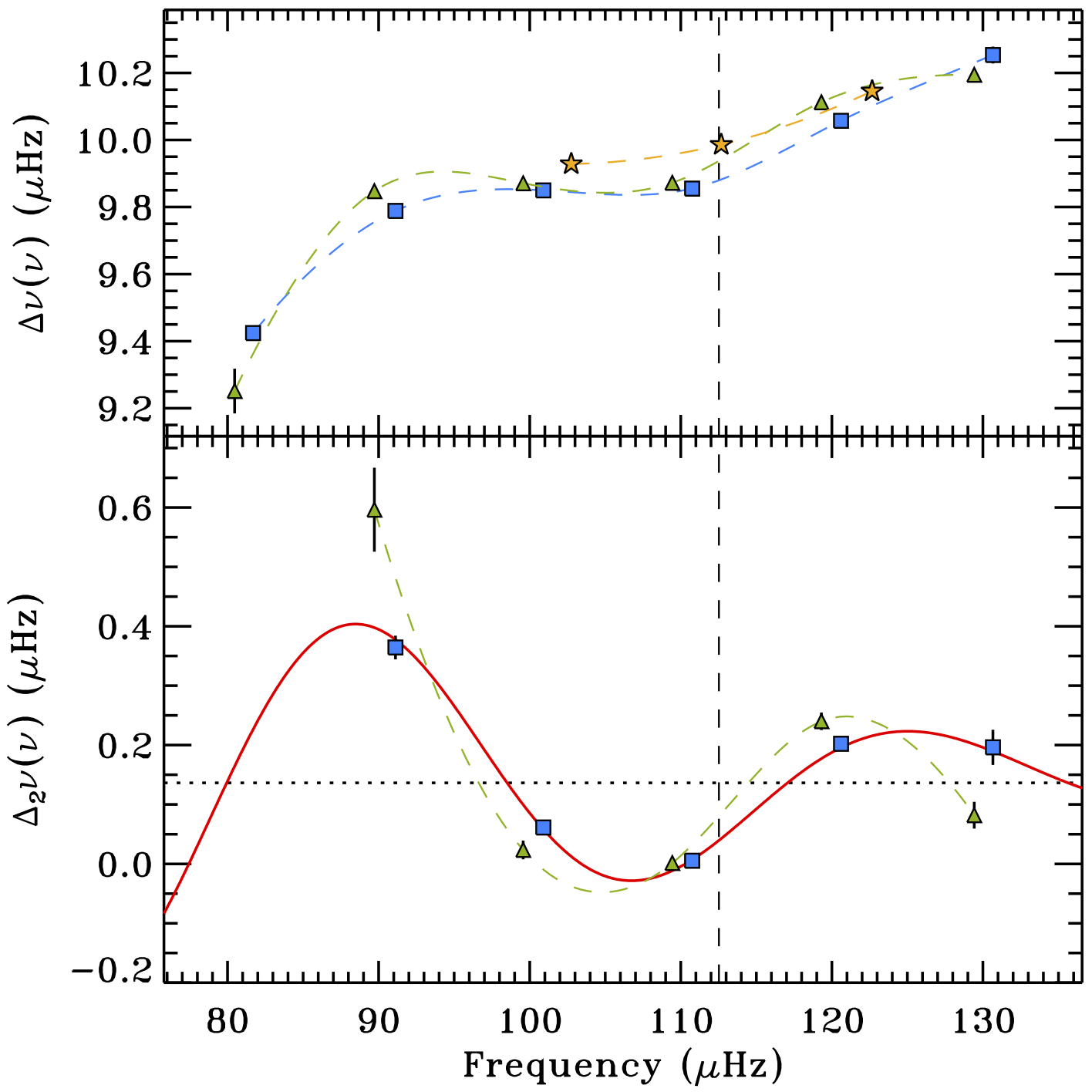}
      \caption{Same description as in Fig.~\ref{fig:glitch} but for KIC~3744043, with yellow star from $\ell = 3$ modes and corresponding polynomial fit with same color.}
    \label{fig:3744043glitch}
\end{figure}

\begin{figure}
   \centering
   \includegraphics[width=9.0cm]{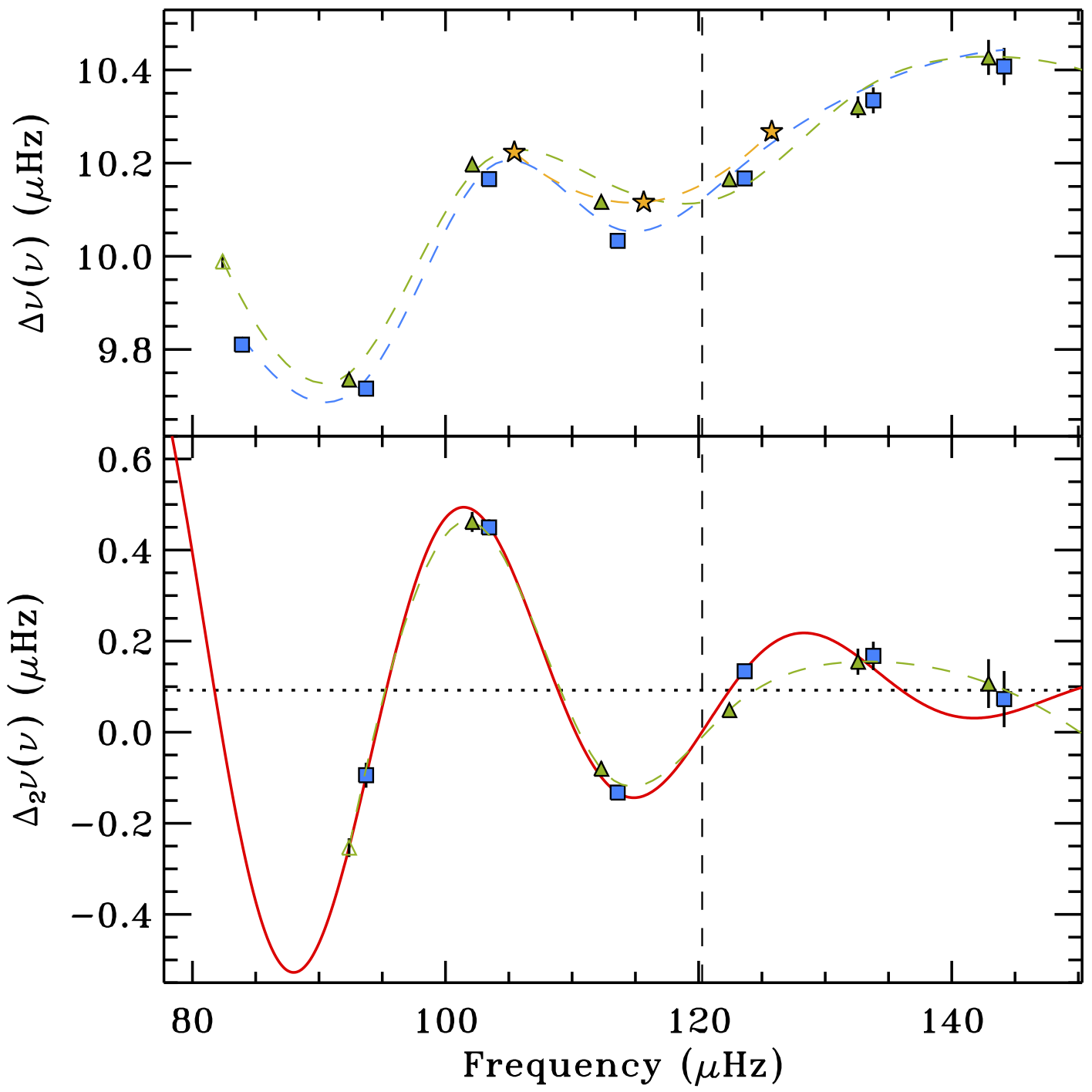}
      \caption{Same description as in Fig.~\ref{fig:glitch} but for KIC~6117517, with yellow star from $\ell = 3$ modes and corresponding polynomial fit with same color. Open symbols represent measurements that used modes with detection probability under the threshold suggested by C15.}
    \label{fig:6117517glitch}
\end{figure}

\begin{figure}
   \centering
   \includegraphics[width=9.0cm]{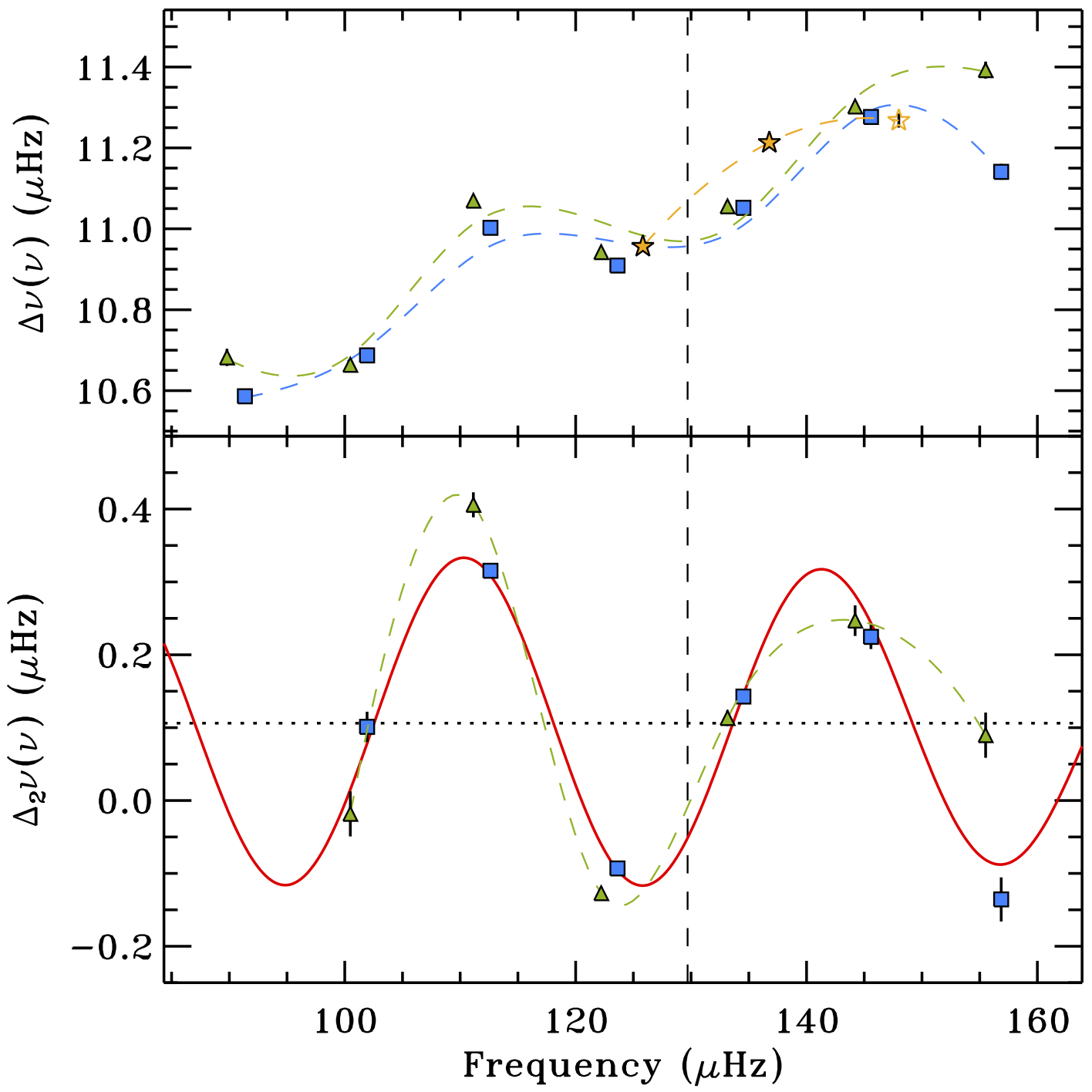}
      \caption{Same description as in Fig.~\ref{fig:glitch} but for KIC~6144777, with yellow star from $\ell = 3$ modes and corresponding polynomial fit with same color.}
    \label{fig:6144777glitch}
\end{figure}

\begin{figure}
   \centering
   \includegraphics[width=9.0cm]{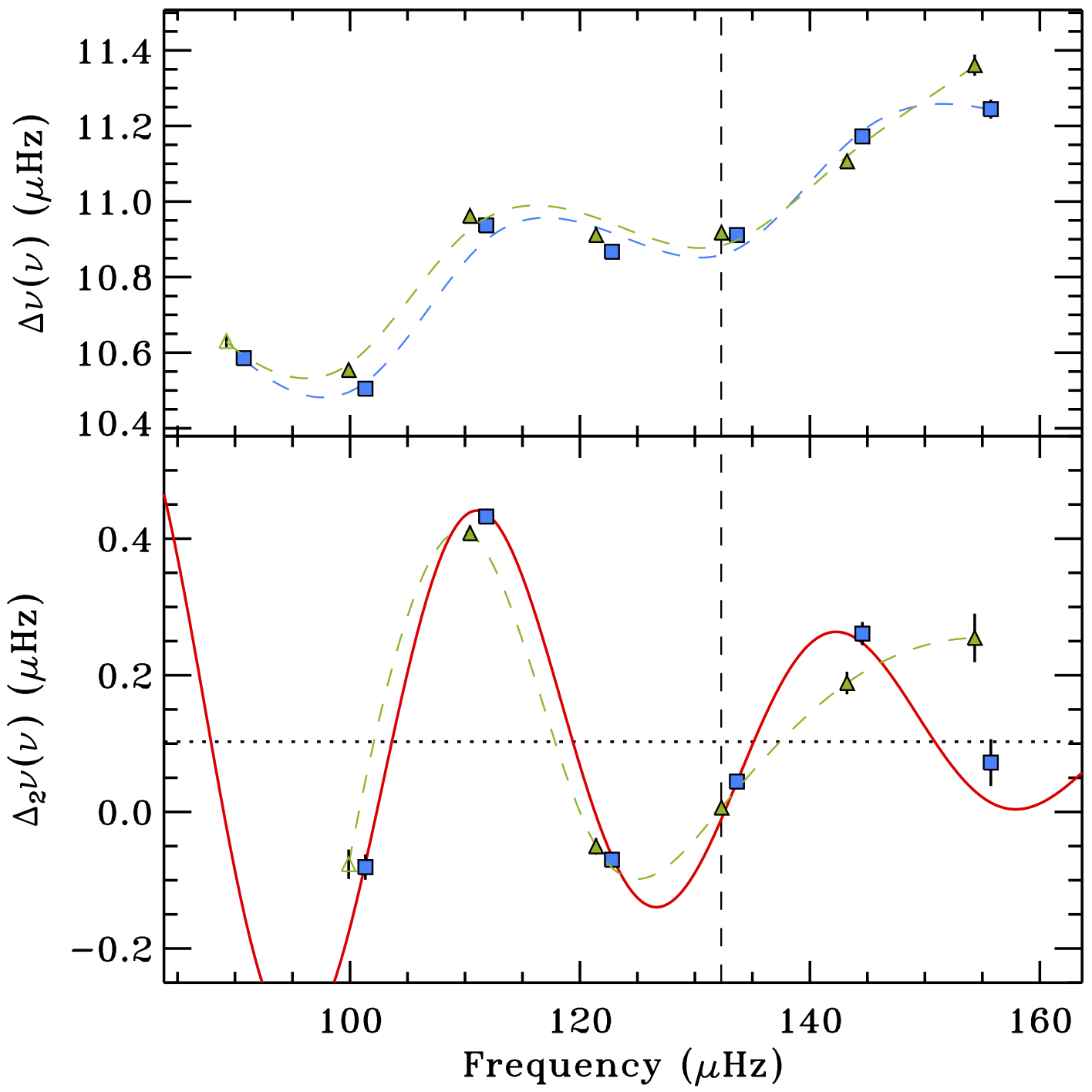}
      \caption{Same description as in Fig.~\ref{fig:glitch} but for KIC~7060732.}
    \label{fig:7060732glitch}
\end{figure}

\begin{figure}
   \centering
   \includegraphics[width=9.0cm]{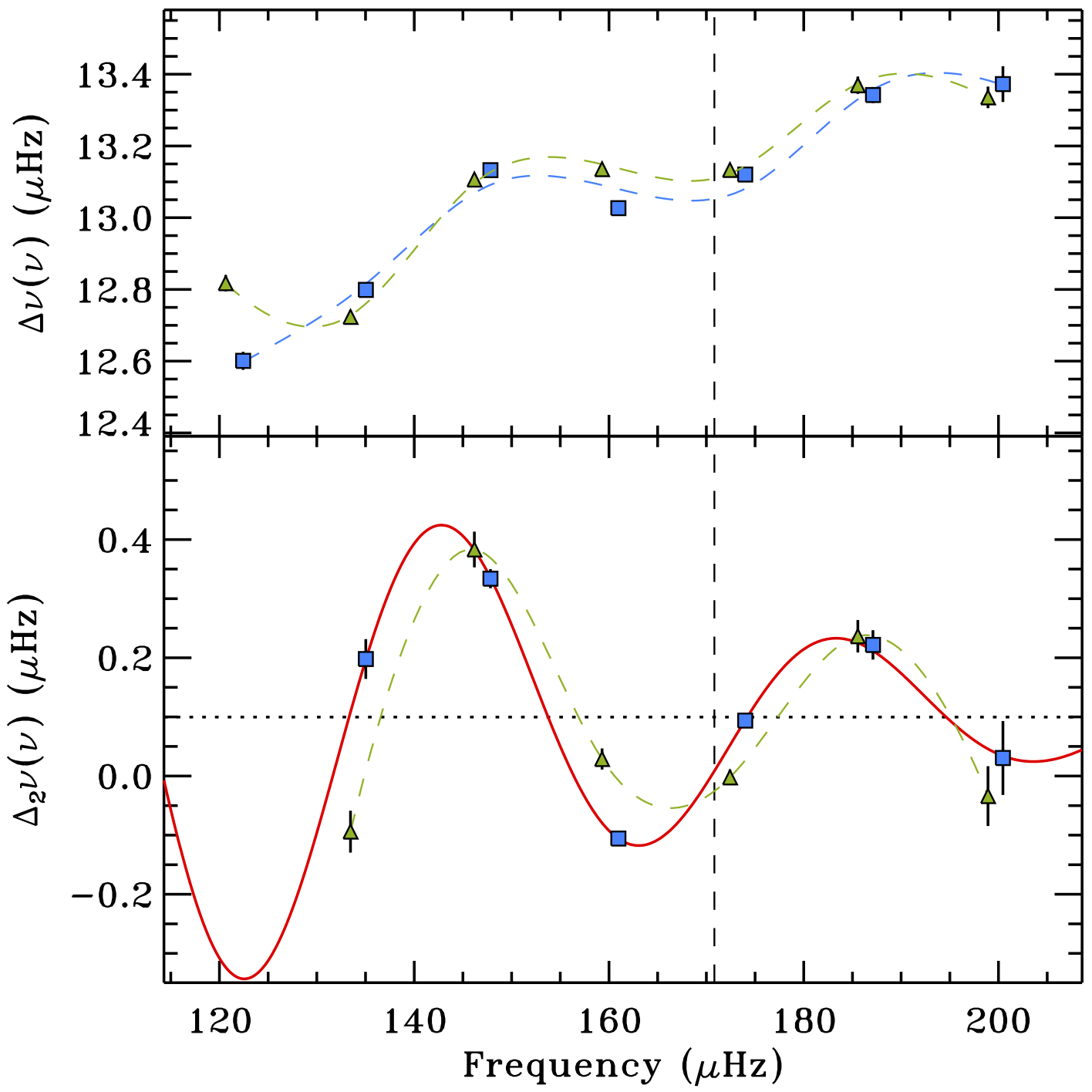}
      \caption{Same description as in Fig.~\ref{fig:glitch} but for KIC~7619745.}
    \label{fig:7619745glitch}
\end{figure}

\begin{figure}
   \centering
   \includegraphics[width=9.0cm]{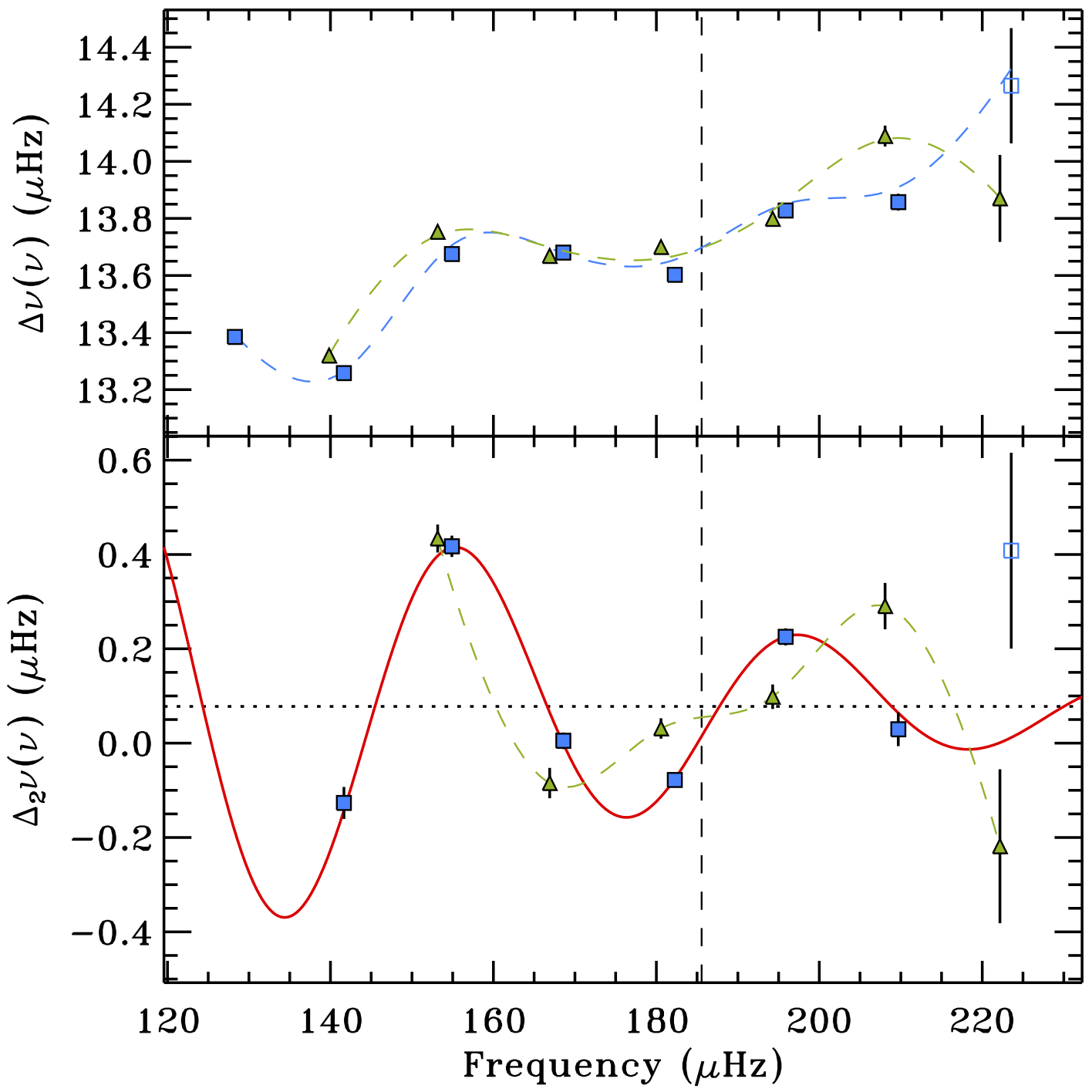}
      \caption{Same description as in Fig.~\ref{fig:glitch} but for KIC~8366239. Open symbols represent measurements that used modes with detection probability under the threshold suggested by C15.}
    \label{fig:8366239glitch}
\end{figure}

\begin{figure}
   \centering
   \includegraphics[width=9.0cm]{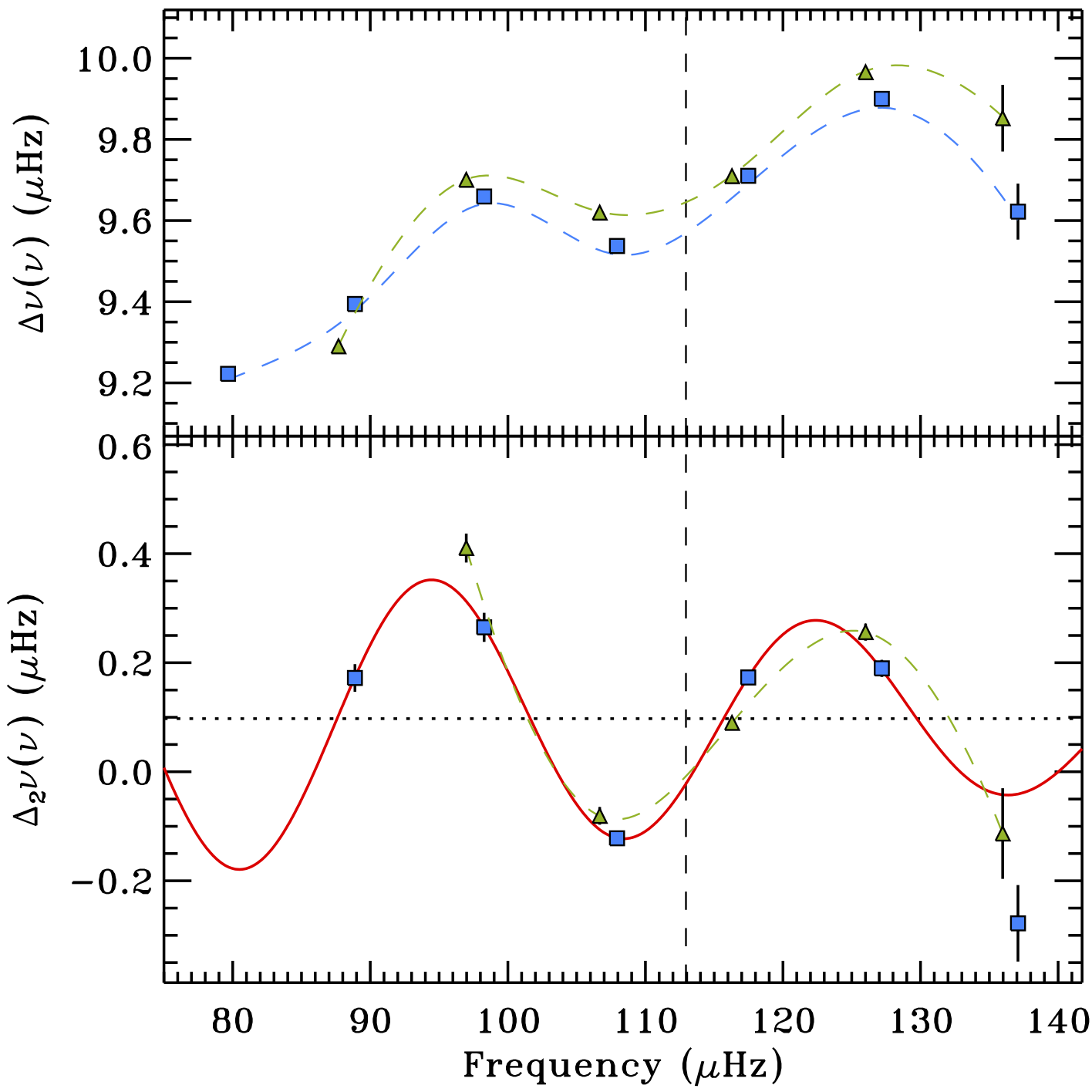}
      \caption{Same description as in Fig.~\ref{fig:glitch} but for KIC~8475025.}
    \label{fig:8475025glitch}
\end{figure}

\begin{figure}
   \centering
   \includegraphics[width=9.0cm]{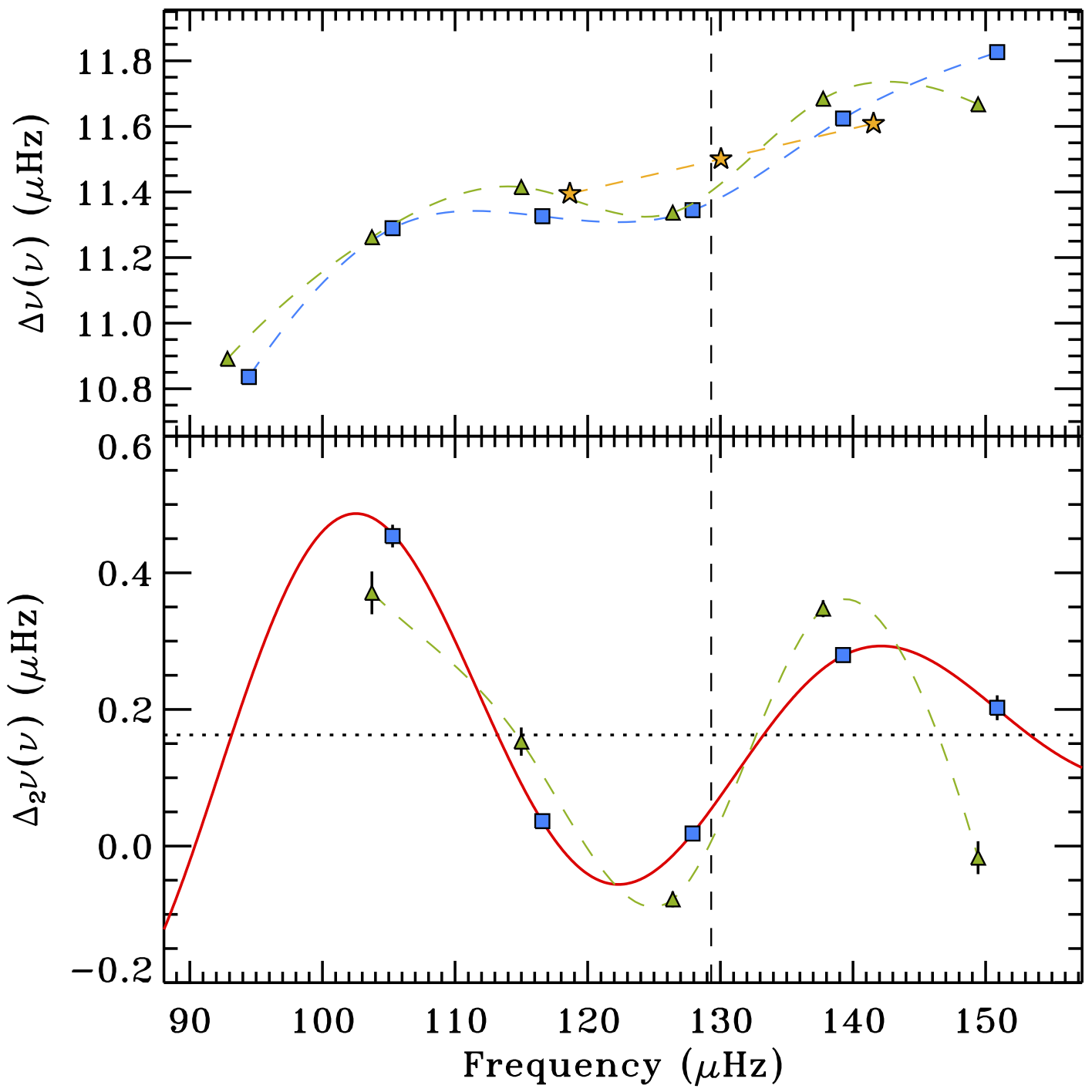}
      \caption{Same description as in Fig.~\ref{fig:glitch} but for KIC~8718745, with yellow star from $\ell = 3$ modes and corresponding polynomial fit with same color.}
    \label{fig:8718745glitch}
\end{figure}

\begin{figure}
   \centering
   \includegraphics[width=9.0cm]{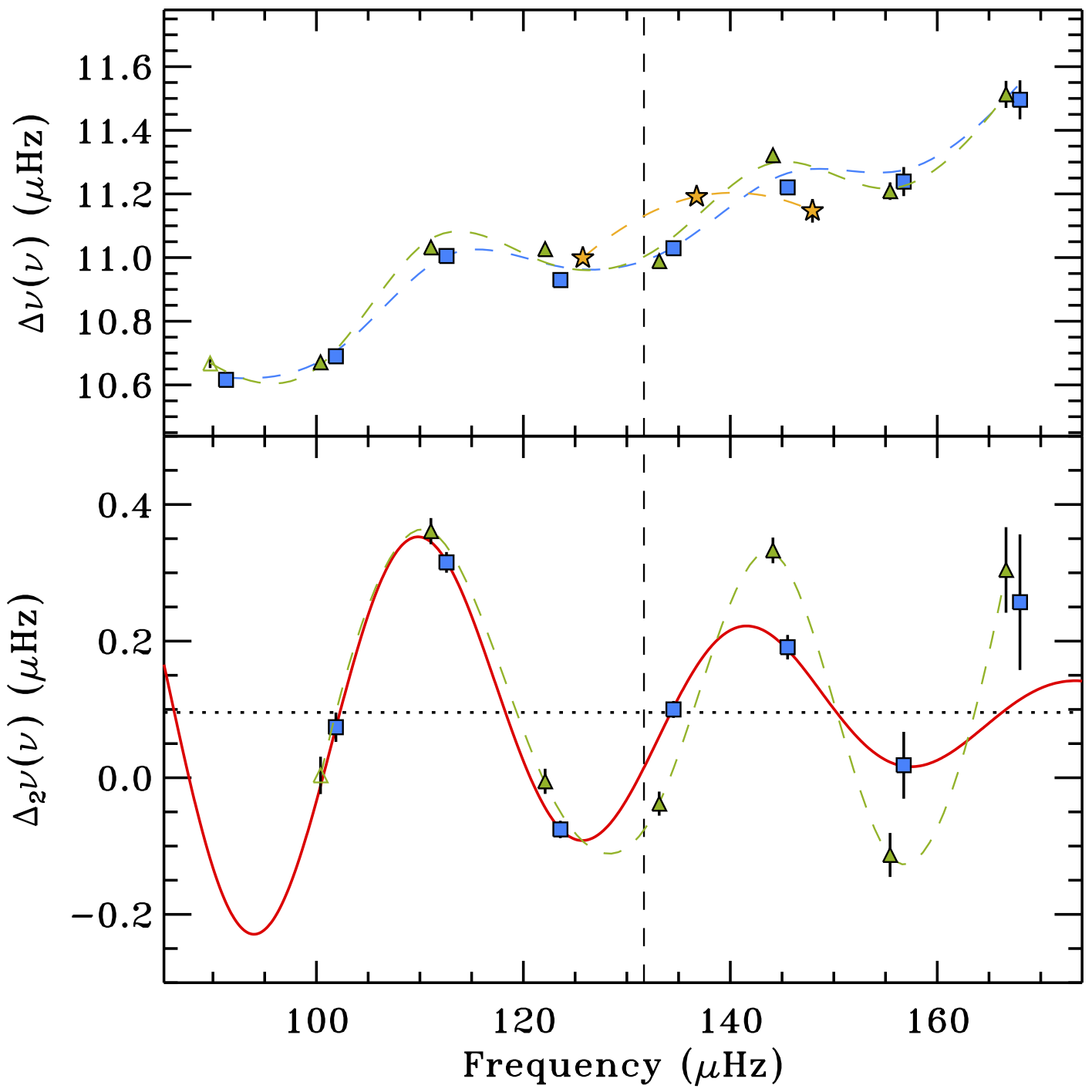}
      \caption{Same description as in Fig.~\ref{fig:glitch} but for KIC~9145955, with yellow star from $\ell = 3$ modes and corresponding polynomial fit with same color. Open symbols represent measurements that used modes with detection probability under the threshold suggested by C15.}
    \label{fig:9145955glitch}
\end{figure}

\begin{figure}
   \centering
   \includegraphics[width=9.0cm]{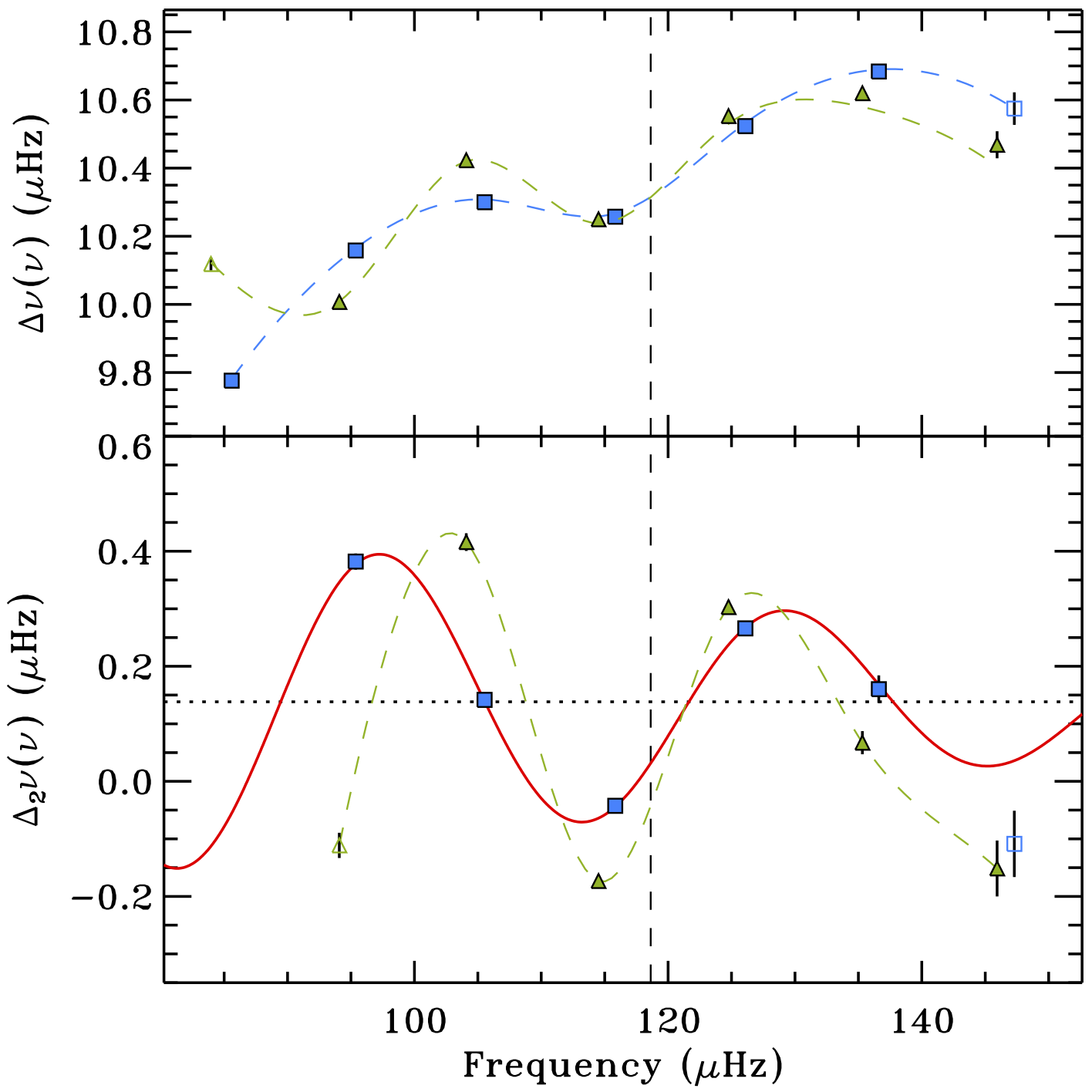}
      \caption{Same description as in Fig.~\ref{fig:glitch} but for KIC~9267654. Open symbols represent measurements that used modes with detection probability under the threshold suggested by C15.}
    \label{fig:9267654glitch}
\end{figure}

\begin{figure}
   \centering
   \includegraphics[width=9.0cm]{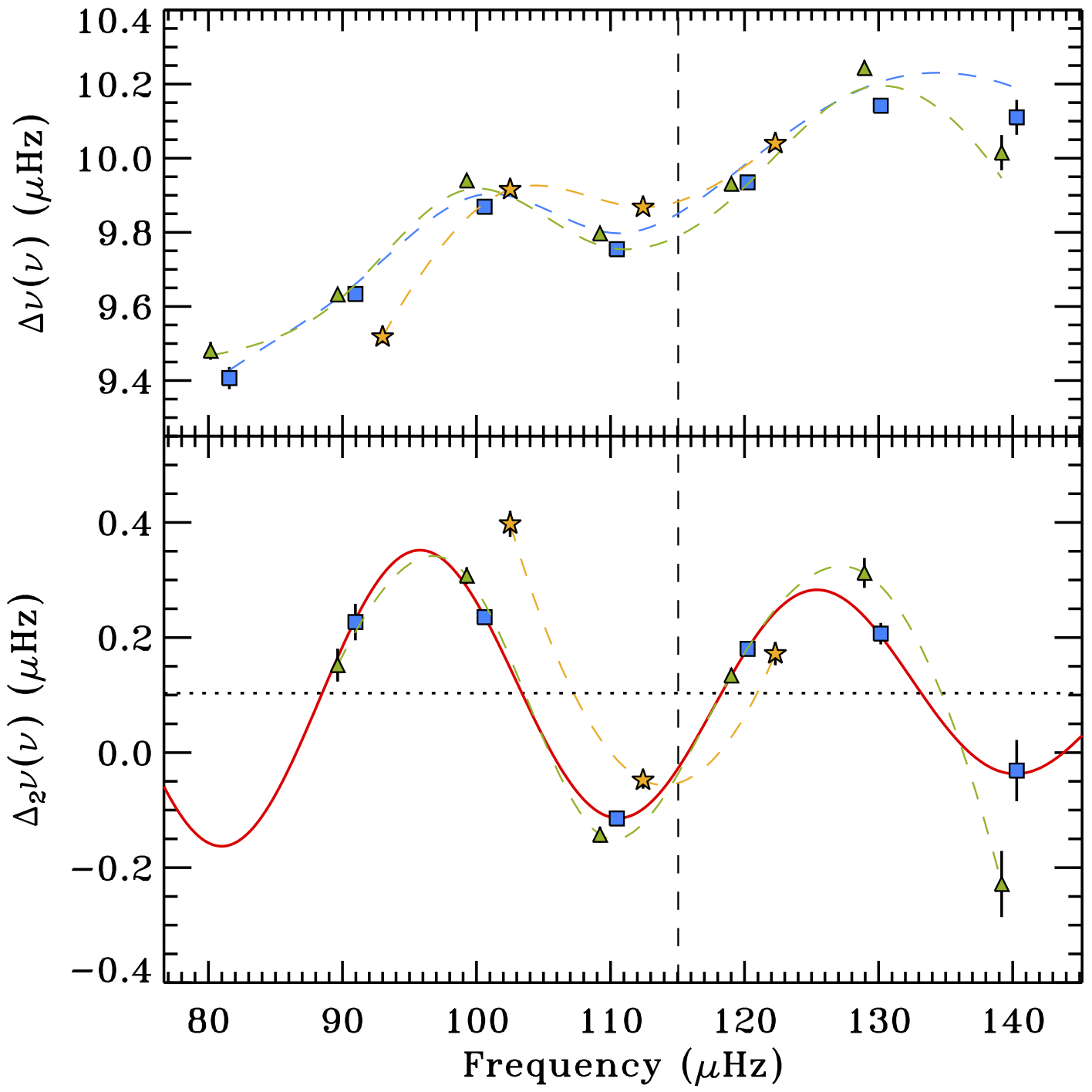}
      \caption{Same description as in Fig.~\ref{fig:glitch} but for KIC~9475697, with yellow star from $\ell = 3$ modes and corresponding polynomial fit with same color. Open symbols represent measurements that used modes with detection probability under the threshold suggested by C15.}
    \label{fig:9475697glitch}
\end{figure}

\begin{figure}
   \centering
   \includegraphics[width=9.0cm]{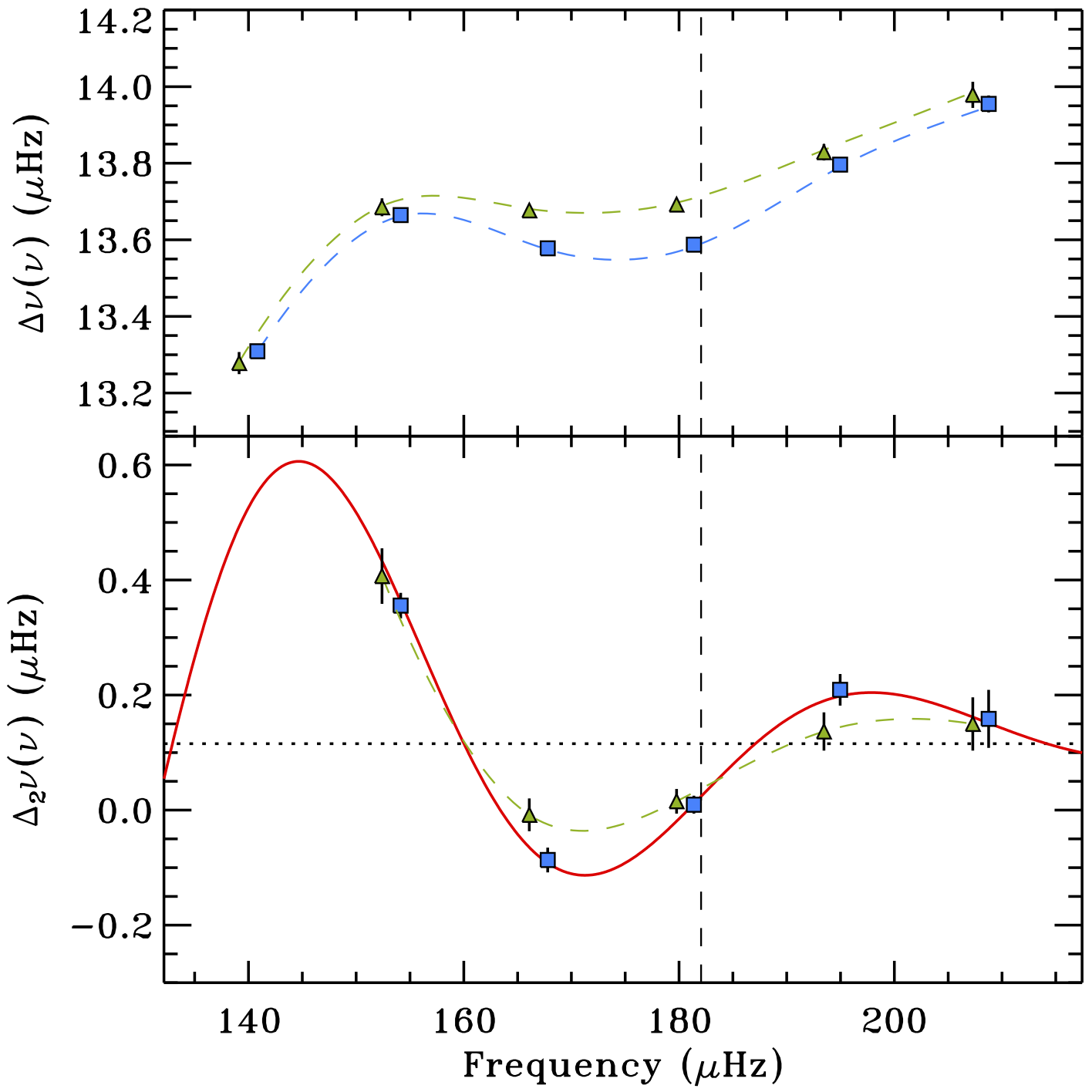}
      \caption{Same description as in Fig.~\ref{fig:glitch} but for KIC~9882316.}
    \label{fig:9882316glitch}
\end{figure}

\begin{figure}
   \centering
   \includegraphics[width=9.0cm]{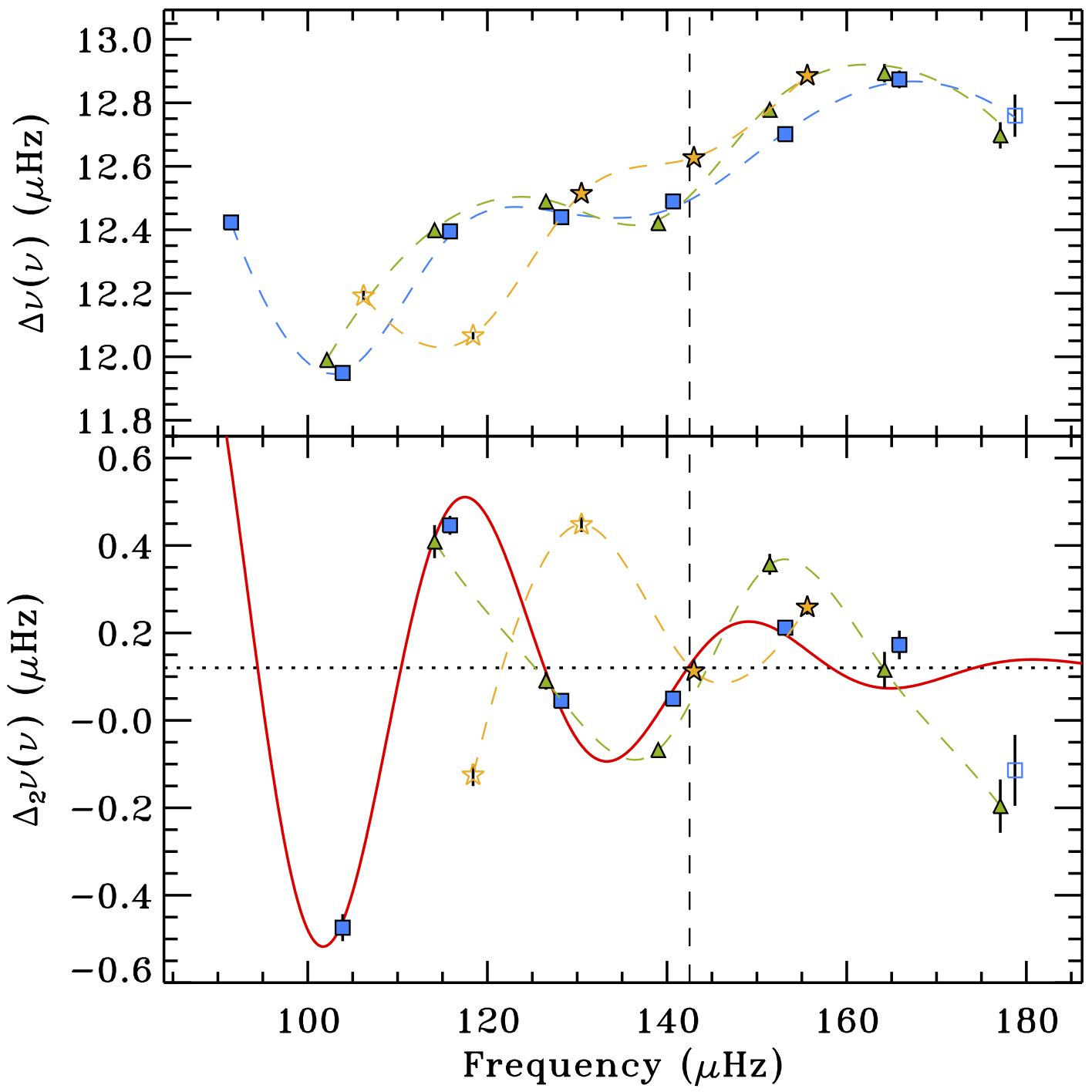}
      \caption{Same description as in Fig.~\ref{fig:glitch} but for KIC~10200377, with yellow star from $\ell = 3$ modes and corresponding polynomial fit with same color. Open symbols represent measurements that used modes with detection probability under the threshold suggested by C15.}
    \label{fig:10200377glitch}
\end{figure}

\begin{figure}
   \centering
   \includegraphics[width=9.0cm]{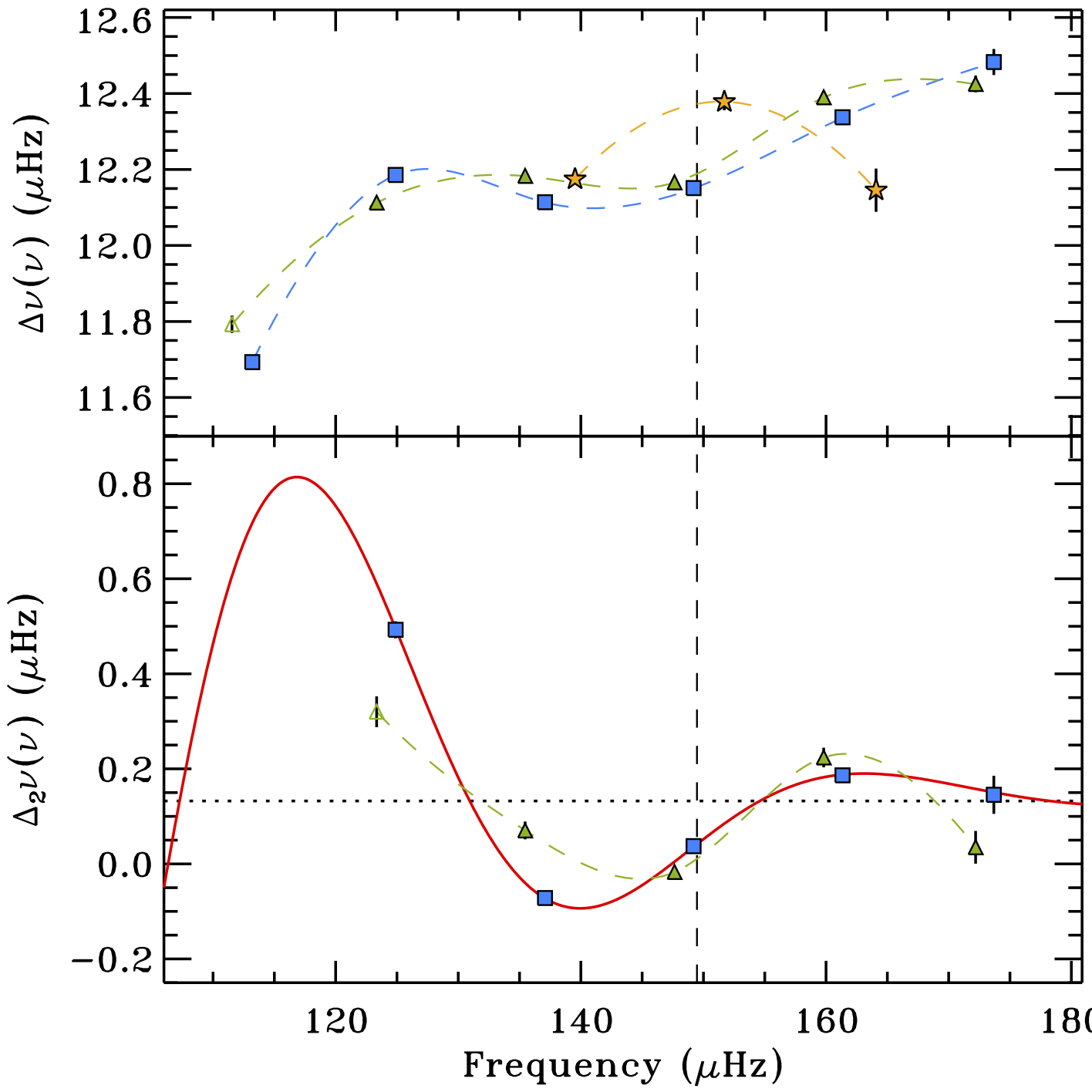}
      \caption{Same description as in Fig.~\ref{fig:glitch} but for KIC~10257278, with yellow star from $\ell = 3$ modes and corresponding polynomial fit with same color. Open symbols represent measurements that used modes with detection probability under the threshold suggested by C15.}
    \label{fig:10257278glitch}
\end{figure}

\begin{figure}
   \centering
   \includegraphics[width=9.0cm]{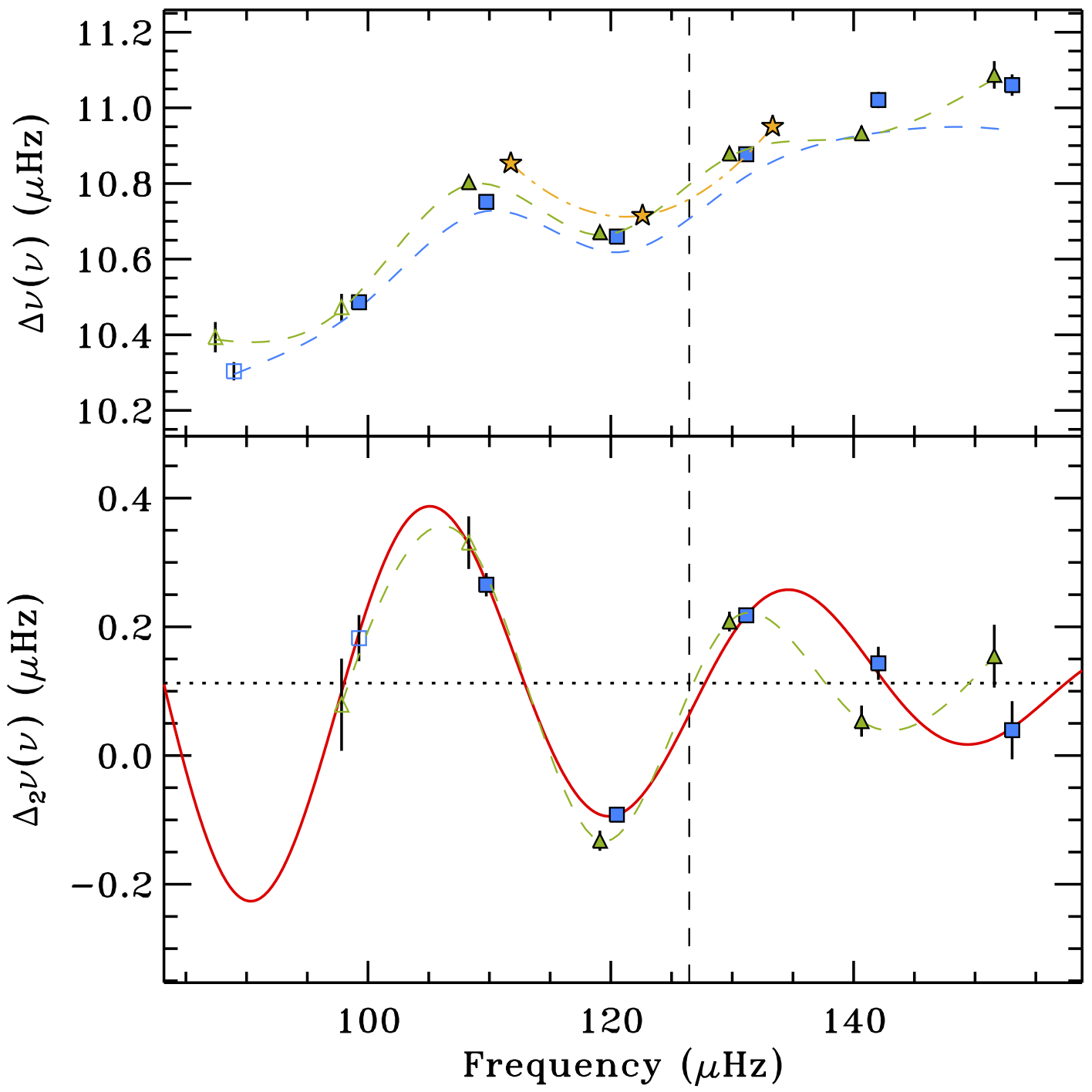}
      \caption{Same description as in Fig.~\ref{fig:glitch} but for KIC~11353313, with yellow star from $\ell = 3$ modes and corresponding polynomial fit with same color. Open symbols represent measurements that used modes with detection probability under the threshold suggested by C15.}
    \label{fig:11353313glitch}
\end{figure}

\begin{figure}
   \centering
   \includegraphics[width=9.0cm]{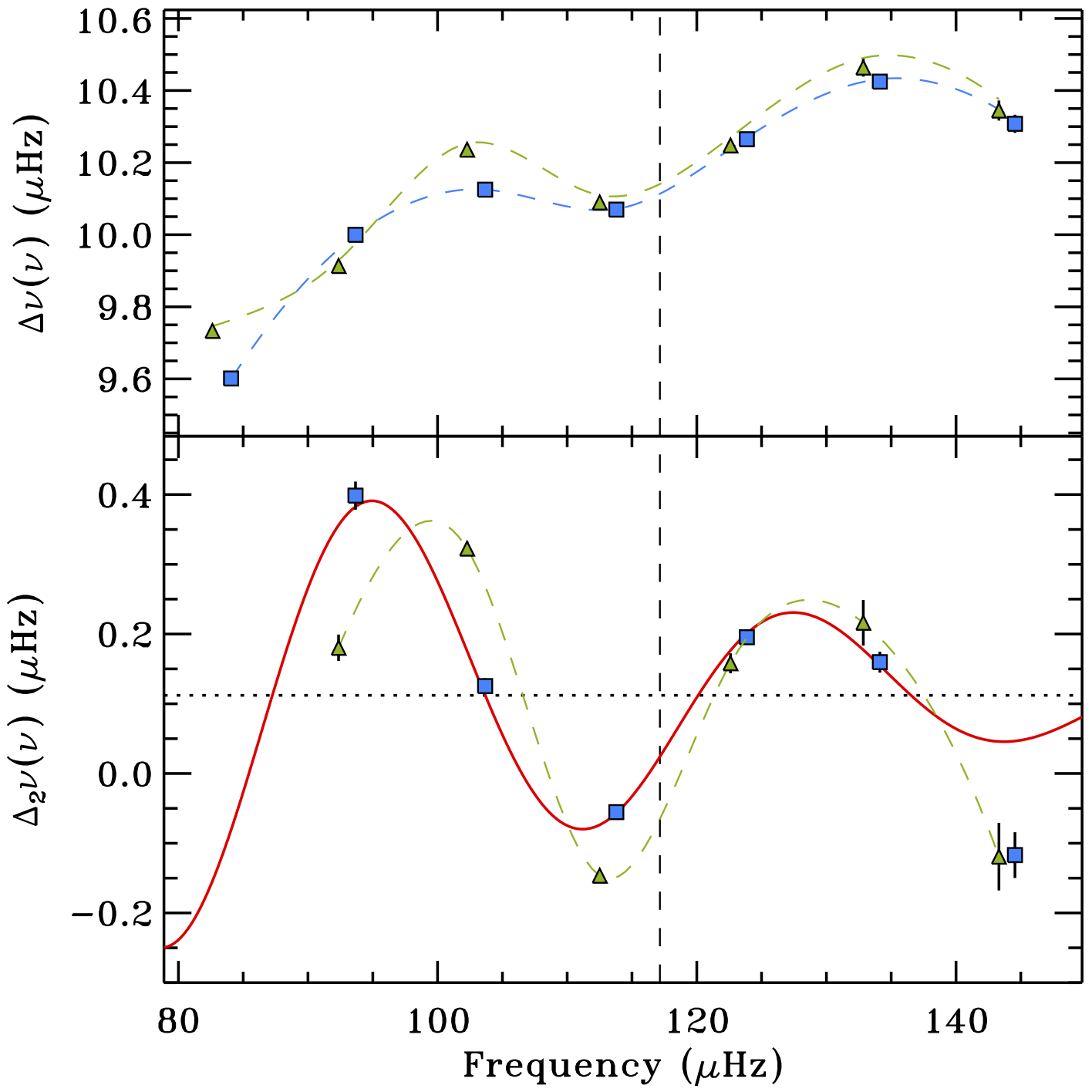}
      \caption{Same description as in Fig.~\ref{fig:glitch} but for KIC~11913545.}
    \label{fig:11913545glitch}
\end{figure}

\begin{figure}
   \centering
   \includegraphics[width=9.0cm]{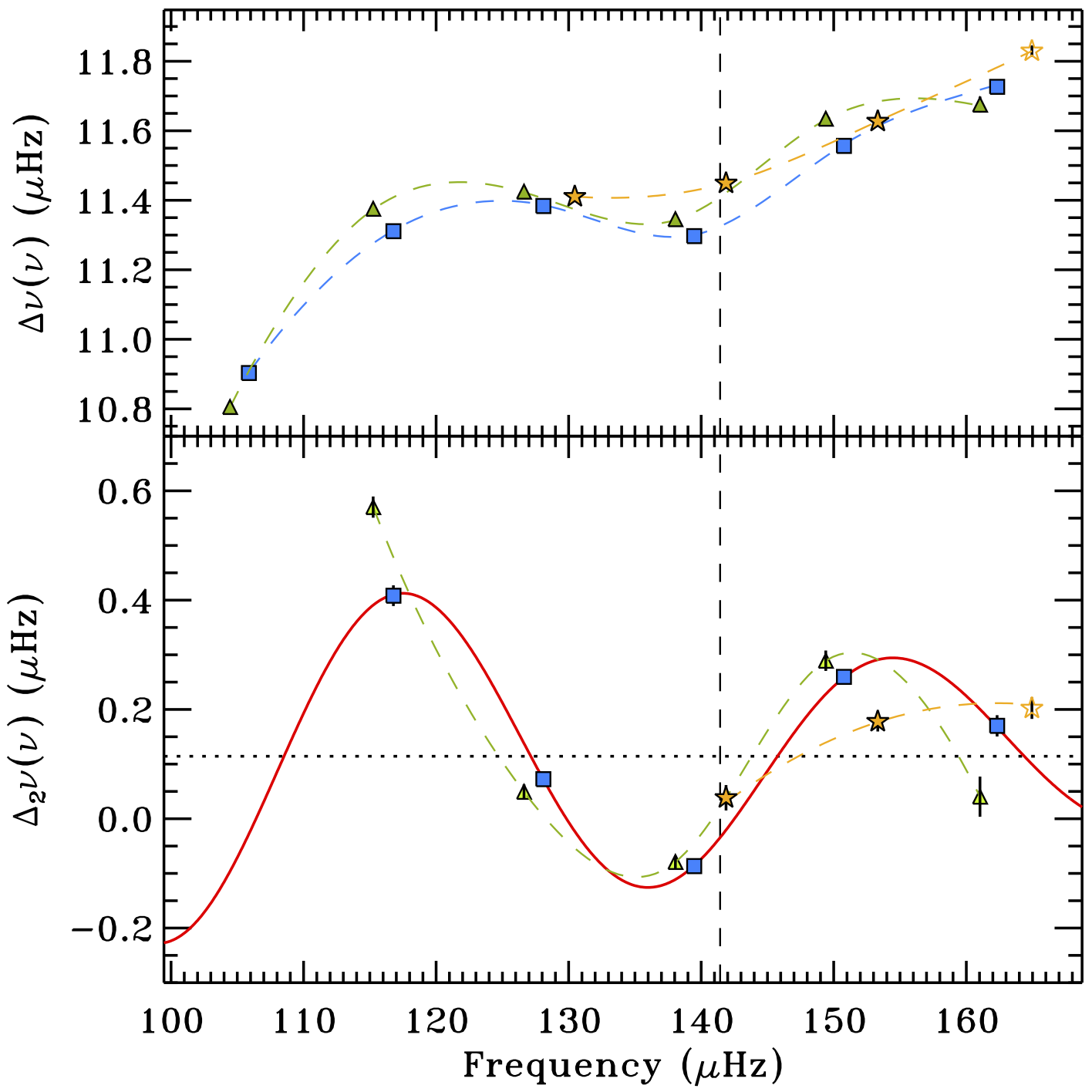}
      \caption{Same description as in Fig.~\ref{fig:glitch} but for KIC~11968334, with yellow star from $\ell = 3$ modes and corresponding polynomial fit with same color. Open symbols represent measurements that used modes with detection probability under the threshold suggested by C15.}
    \label{fig:11968334glitch}
\end{figure}

All the individual angular frequency measurements and corresponding second differences of the radial modes are listed in the Tables \ref{tab:2diff} and \ref{tab:2diff_2}. The results for the first differences $\Dnunu$ and second differences $\Dnunua$ of all the LRGs, are shown in the Figs. from \ref{fig:3744043glitch} to \ref{fig:11968334glitch}, similarly to that provided for \kic\,in Fig.~\ref{fig:glitch}.
\label{sec:glitch_results}
% Figures of acoustic glitches for all the stars

\begin{table}
\caption{Median values with corresponding 68.3\,\% shortest credible intervals for the radial angular frequencies $\omega_{n,0}$ and corresponding second angular frequency differences $\Delta_2 \omega_{n,0}$ with 1-$\sigma$ error bars, as described in Sect.~\ref{sec:analysis}, for all the stars of the sample.}             % title of Table
\centering                         
\begin{tabular}{l l r c}       
\hline\hline
\\[-8pt]
KIC ID & \multicolumn{1}{c}{$\omega_{n,0}$} & \multicolumn{1}{c}{$\Delta_2 \omega_{n,0}$} \\ [1pt]
& \multicolumn{1}{c}{(\muhz)} & \multicolumn{1}{c}{(\muhz)} \\ [1pt]
\hline
\\[-8pt]
03744043 & $      572.48_{-        0.05}^{+        0.05}$ & $        2.29_{-        0.13}^{+        0.13}$ \\[1pt]
 & $      633.98_{-        0.03}^{+        0.03}$ & $        0.39_{-        0.08}^{+        0.08}$ \\[1pt]
 & $      695.87_{-        0.02}^{+        0.01}$ & $        0.03_{-        0.05}^{+        0.05}$ \\[1pt]
 & $      757.79_{-        0.03}^{+        0.02}$ & $        1.27_{-        0.07}^{+        0.07}$ \\[1pt]
 & $      820.98_{-        0.05}^{+        0.06}$ & $        1.23_{-        0.19}^{+        0.19}$ \\[1pt]
 \hline
\\[-8pt]
06117517  & $      588.87_{-        0.07}^{+        0.08}$ & $       -0.59_{-        0.17}^{+        0.17}$ \\[1pt]
 & $      649.91_{-        0.04}^{+        0.04}$ & $        2.83_{-        0.11}^{+        0.11}$ \\[1pt]
 & $      713.79_{-        0.01}^{+        0.01}$ & $       -0.83_{-        0.05}^{+        0.05}$ \\[1pt]
 & $      776.83_{-        0.02}^{+        0.02}$ & $        0.84_{-        0.06}^{+        0.06}$ \\[1pt]
 & $      840.71_{-        0.05}^{+        0.05}$ & $        1.05_{-        0.20}^{+        0.20}$ \\[1pt]
 & $      905.64_{-        0.16}^{+        0.17}$ & $        0.46_{-        0.39}^{+        0.39}$ \\[1pt]
 \hline
\\[-8pt]
06144777 & $      640.47_{-        0.05}^{+        0.05}$ & $        0.64_{-        0.13}^{+        0.13}$ \\[1pt]
& $      707.62_{-        0.02}^{+        0.02}$ & $        1.98_{-        0.06}^{+        0.06}$ \\[1pt]
&  $      776.75_{-        0.01}^{+        0.01}$ & $       -0.59_{-        0.04}^{+        0.04}$ \\[1pt]
&  $      845.29_{-        0.01}^{+        0.01}$ & $        0.90_{-        0.04}^{+        0.04}$ \\[1pt]
&  $      914.74_{-        0.03}^{+        0.03}$ & $        1.41_{-        0.10}^{+        0.10}$ \\[1pt]
&  $      985.59_{-        0.08}^{+        0.08}$ & $       -0.85_{-        0.19}^{+        0.19}$ \\[1pt]
\hline
\\[-8pt]
07060732 & $      636.76_{-        0.05}^{+        0.05}$ & $       -0.51_{-        0.12}^{+        0.12}$ \\[1pt]
&  $      702.76_{-        0.02}^{+        0.02}$ & $        2.72_{-        0.07}^{+        0.07}$ \\[1pt]
&  $      771.48_{-        0.03}^{+        0.03}$ & $       -0.44_{-        0.07}^{+        0.07}$ \\[1pt]
&  $      839.76_{-        0.02}^{+        0.02}$ & $        0.28_{-        0.06}^{+        0.06}$ \\[1pt]
&  $      908.32_{-        0.03}^{+        0.03}$ & $        1.64_{-        0.11}^{+        0.11}$ \\[1pt]
&  $      978.51_{-        0.09}^{+        0.08}$ & $        0.45_{-        0.22}^{+        0.22}$ \\[1pt]      
\hline
\\[-8pt]
07619745 & $      848.43_{-        0.08}^{+        0.08}$ & $        1.24_{-        0.21}^{+        0.21}$ \\[1pt]
 & $      928.84_{-        0.03}^{+        0.03}$ & $        2.10_{-        0.10}^{+        0.10}$ \\[1pt]
 & $     1011.36_{-        0.02}^{+        0.02}$ & $       -0.66_{-        0.05}^{+        0.05}$ \\[1pt]
 & $     1093.21_{-        0.02}^{+        0.02}$ & $        0.59_{-        0.06}^{+        0.06}$ \\[1pt]
 & $     1175.64_{-        0.04}^{+        0.04}$ & $        1.39_{-        0.16}^{+        0.16}$ \\[1pt]
 & $     1259.48_{-        0.14}^{+        0.13}$ & $        0.19_{-        0.39}^{+        0.39}$ \\[1pt]           
 \hline
\\[-8pt]
08366239 & $      890.01_{-        0.09}^{+        0.09}$ & $       -0.80_{-        0.21}^{+        0.21}$ \\[1pt]
&  $      973.31_{-        0.05}^{+        0.05}$ & $        2.62_{-        0.14}^{+        0.14}$ \\[1pt]
&  $     1059.23_{-        0.05}^{+        0.05}$ & $        0.03_{-        0.11}^{+        0.11}$ \\[1pt]
&  $     1145.19_{-        0.02}^{+        0.02}$ & $       -0.49_{-        0.08}^{+        0.08}$ \\[1pt]
&  $     1230.66_{-        0.04}^{+        0.05}$ & $        1.42_{-        0.11}^{+        0.11}$ \\[1pt]
&  $     1317.54_{-        0.07}^{+        0.07}$ & $        0.18_{-        0.22}^{+        0.22}$ \\[1pt]
&  $     1404.60_{-        0.18}^{+        0.17}$ & $        2.56_{-        1.30}^{+        1.30}$ \\[1pt]         
\hline
\\[-8pt]
08475025 & $      558.46_{-        0.06}^{+        0.07}$ & $        1.08_{-        0.16}^{+        0.16}$ \\[1pt]
&  $      617.49_{-        0.08}^{+        0.08}$ & $        1.67_{-        0.17}^{+        0.17}$ \\[1pt]
&  $      678.18_{-        0.02}^{+        0.02}$ & $       -0.77_{-        0.09}^{+        0.09}$ \\[1pt]
&  $      738.10_{-        0.02}^{+        0.02}$ & $        1.09_{-        0.06}^{+        0.06}$ \\[1pt]
&  $      799.12_{-        0.04}^{+        0.04}$ & $        1.19_{-        0.10}^{+        0.10}$ \\[1pt]
&  $      861.32_{-        0.04}^{+        0.04}$ & $       -1.75_{-        0.44}^{+        0.44}$ \\[1pt] 
\hline
\\[-8pt]
08718745 & $      661.54_{-        0.04}^{+        0.04}$ & $        2.85_{-        0.10}^{+        0.10}$ \\[1pt]
& $      732.48_{-        0.03}^{+        0.03}$ & $        0.23_{-        0.07}^{+        0.07}$ \\[1pt]
& $      803.64_{-        0.01}^{+        0.01}$ & $        0.12_{-        0.05}^{+        0.05}$ \\[1pt]
& $      874.92_{-        0.02}^{+        0.03}$ & $        1.76_{-        0.07}^{+        0.07}$ \\[1pt]
& $      947.96_{-        0.04}^{+        0.04}$ & $        1.27_{-        0.11}^{+        0.11}$ \\[1pt]    
\hline
\\[-8pt]
09145955 & $      640.13_{-        0.05}^{+        0.06}$ & $        0.46_{-        0.14}^{+        0.14}$ \\[1pt]
 & $      707.29_{-        0.03}^{+        0.04}$ & $        1.98_{-        0.09}^{+        0.09}$ \\[1pt]
 & $      776.44_{-        0.04}^{+        0.03}$ & $       -0.48_{-        0.08}^{+        0.08}$ \\[1pt]
 & $      845.11_{-        0.03}^{+        0.03}$ & $        0.63_{-        0.08}^{+        0.08}$ \\[1pt]
 & $      914.41_{-        0.05}^{+        0.04}$ & $        1.20_{-        0.11}^{+        0.11}$ \\[1pt]
 & $      984.91_{-        0.06}^{+        0.06}$ & $        0.11_{-        0.31}^{+        0.31}$ \\[1pt]
 & $     1055.53_{-        0.28}^{+        0.29}$ & $        1.61_{-        0.62}^{+        0.62}$ \\[1pt]
 \hline    
\end{tabular}
\label{tab:2diff}
\end{table}

\begin{table}
\caption{Table~\ref{tab:2diff} continued.}             % title of Table
\centering                         
\begin{tabular}{l l r c}       
\hline\hline
\\[-8pt]
KIC ID & \multicolumn{1}{c}{$\omega_{n,0}$} & \multicolumn{1}{c}{$\Delta_2 \omega_{n,0}$} \\ [1pt]
& \multicolumn{1}{c}{(\muhz)} & \multicolumn{1}{c}{(\muhz)} \\ [1pt]
\hline
\\[-8pt]
09267654 & $      599.25_{-        0.03}^{+        0.03}$ & $        2.40_{-        0.09}^{+        0.09}$ \\[1pt]
&  $      663.07_{-        0.03}^{+        0.03}$ & $        0.89_{-        0.07}^{+        0.07}$ \\[1pt]
&  $      727.79_{-        0.01}^{+        0.01}$ & $       -0.27_{-        0.04}^{+        0.04}$ \\[1pt]
&  $      792.24_{-        0.02}^{+        0.02}$ & $        1.67_{-        0.06}^{+        0.06}$ \\[1pt]
&  $      858.36_{-        0.05}^{+        0.05}$ & $        1.01_{-        0.15}^{+        0.15}$ \\[1pt]
&  $      925.49_{-        0.12}^{+        0.11}$ & $       -0.68_{-        0.36}^{+        0.36}$ \\[1pt]
\hline
\\[-8pt]
09475697 & $      571.57_{-        0.04}^{+        0.03}$ & $        1.43_{-        0.20}^{+        0.20}$ \\[1pt]
&  $      632.10_{-        0.03}^{+        0.03}$ & $        1.48_{-        0.07}^{+        0.07}$ \\[1pt]
&  $      694.11_{-        0.02}^{+        0.02}$ & $       -0.72_{-        0.05}^{+        0.05}$ \\[1pt]
&  $      755.40_{-        0.01}^{+        0.01}$ & $        1.13_{-        0.05}^{+        0.05}$ \\[1pt]
&  $      817.82_{-        0.04}^{+        0.04}$ & $        1.30_{-        0.12}^{+        0.12}$ \\[1pt]
&  $      881.54_{-        0.09}^{+        0.09}$ & $       -0.20_{-        0.33}^{+        0.33}$ \\[1pt]
\hline
\\[-8pt]
09882316 & $      968.46_{-        0.05}^{+        0.05}$ & $        2.23_{-        0.14}^{+        0.14}$ \\[1pt]
&  $     1054.32_{-        0.06}^{+        0.06}$ & $       -0.54_{-        0.13}^{+        0.13}$ \\[1pt]
&  $     1139.63_{-        0.03}^{+        0.03}$ & $        0.06_{-        0.10}^{+        0.10}$ \\[1pt]
&  $     1225.00_{-        0.05}^{+        0.05}$ & $        1.31_{-        0.17}^{+        0.17}$ \\[1pt]
&  $     1311.69_{-        0.14}^{+        0.14}$ & $        1.00_{-        0.32}^{+        0.32}$ \\[1pt]
\hline
\\[-8pt]
10200377 & $      652.74_{-        0.09}^{+        0.09}$ & $       -2.98_{-        0.19}^{+        0.19}$ \\[1pt]
&  $      727.81_{-        0.05}^{+        0.05}$ & $        2.80_{-        0.14}^{+        0.14}$ \\[1pt]
&  $      805.69_{-        0.03}^{+        0.03}$ & $        0.28_{-        0.08}^{+        0.08}$ \\[1pt]
&  $      883.85_{-        0.02}^{+        0.02}$ & $        0.31_{-        0.05}^{+        0.05}$ \\[1pt]
&  $      962.33_{-        0.02}^{+        0.02}$ & $        1.33_{-        0.08}^{+        0.08}$ \\[1pt]
&  $     1042.13_{-        0.06}^{+        0.06}$ & $        1.08_{-        0.21}^{+        0.21}$ \\[1pt]
&  $     1123.02_{-        0.17}^{+        0.16}$ & $       -0.72_{-        0.51}^{+        0.51}$ \\[1pt]
\hline
\\[-8pt]
10257278 & $      784.72_{-        0.05}^{+        0.05}$ & $        3.10_{-        0.11}^{+        0.11}$ \\[1pt]
& $      861.29_{-        0.02}^{+        0.02}$ & $       -0.45_{-        0.07}^{+        0.07}$ \\[1pt]
& $      937.40_{-        0.02}^{+        0.02}$ & $        0.23_{-        0.05}^{+        0.05}$ \\[1pt]
& $     1013.75_{-        0.02}^{+        0.03}$ & $        1.17_{-        0.09}^{+        0.09}$ \\[1pt]
& $     1091.27_{-        0.07}^{+        0.08}$ & $        0.91_{-        0.25}^{+        0.25}$ \\[1pt]
\hline
\\[-8pt]
11353313 & $      623.65_{-        0.09}^{+        0.10}$ & $        1.15_{-        0.22}^{+        0.22}$ \\[1pt]
 & $      689.53_{-        0.03}^{+        0.03}$ & $        1.67_{-        0.11}^{+        0.11}$ \\[1pt]
 & $      757.09_{-        0.02}^{+        0.02}$ & $       -0.58_{-        0.05}^{+        0.05}$ \\[1pt]
 & $      824.06_{-        0.02}^{+        0.02}$ & $        1.37_{-        0.07}^{+        0.07}$ \\[1pt]
 & $      892.41_{-        0.05}^{+        0.05}$ & $        0.90_{-        0.16}^{+        0.16}$ \\[1pt]
 & $      961.65_{-        0.13}^{+        0.12}$ & $        0.25_{-        0.28}^{+        0.28}$ \\[1pt]
 \hline
\\[-8pt]
11913545 & $      588.52_{-        0.06}^{+        0.06}$ & $        2.50_{-        0.13}^{+        0.13}$ \\[1pt]
& $      651.35_{-        0.02}^{+        0.02}$ & $        0.79_{-        0.07}^{+        0.07}$ \\[1pt]
& $      714.97_{-        0.01}^{+        0.01}$ & $       -0.35_{-        0.03}^{+        0.03}$ \\[1pt]
& $      778.24_{-        0.01}^{+        0.01}$ & $        1.23_{-        0.04}^{+        0.04}$ \\[1pt]
& $      842.74_{-        0.03}^{+        0.03}$ & $        1.00_{-        0.09}^{+        0.09}$ \\[1pt]
& $      908.24_{-        0.07}^{+        0.08}$ & $       -0.74_{-        0.21}^{+        0.21}$ \\[1pt]
\hline
\\[-8pt]
11968334 & $      733.74_{-        0.04}^{+        0.04}$ & $        2.56_{-        0.12}^{+        0.12}$ \\[1pt]
& $      804.81_{-        0.03}^{+        0.03}$ & $        0.45_{-        0.08}^{+        0.08}$ \\[1pt]
& $      876.34_{-        0.02}^{+        0.01}$ & $       -0.54_{-        0.05}^{+        0.05}$ \\[1pt]
& $      947.32_{-        0.03}^{+        0.03}$ & $        1.63_{-        0.07}^{+        0.07}$ \\[1pt]
& $     1019.93_{-        0.04}^{+        0.03}$ & $        1.07_{-        0.12}^{+        0.12}$ \\[1pt]
\hline
\\[-8pt]
12008916 & $      832.03_{-        0.03}^{+        0.03}$ & $        2.95_{-        0.14}^{+        0.14}$ \\[1pt]
& $      912.25_{-        0.04}^{+        0.04}$ & $        0.69_{-        0.09}^{+        0.09}$ \\[1pt]
& $      993.16_{-        0.02}^{+        0.02}$ & $       -0.74_{-        0.06}^{+        0.06}$ \\[1pt]
& $     1073.33_{-        0.02}^{+        0.03}$ & $        1.75_{-        0.07}^{+        0.07}$ \\[1pt]
& $     1155.25_{-        0.05}^{+        0.05}$ & $        0.95_{-        0.14}^{+        0.14}$ \\[1pt]
\hline    
\end{tabular}
\label{tab:2diff_2}
\end{table}

\end{document}